\documentclass[draft,tightenlines,nofootinbib,preprint,aps,eqsecnum]{revtex4}

\newcommand{\beq}{\begin{equation}}
\newcommand{\eeq}{\end{equation}}
\newcommand{\bea}{\begin{eqnarray}}
\newcommand{\eea}{\end{eqnarray}}
\newcommand{\cir}{{\buildrel \circ \over =}}
\newcommand{\sgn}{\epsilon}
\newcommand{\eo}{{}^4{\buildrel \circ \over E}}

\begin{document}

\title{Canonical ADM Tetrad Gravity: from Metrological Inertial Gauge Variables to
Dynamical Tidal Dirac Observables}

\medskip

\author{Luca Lusanna}

\affiliation{ Sezione INFN di Firenze\\ Polo Scientifico\\ Via Sansone 1\\
50019 Sesto Fiorentino (FI), Italy\\ Phone: 0039-055-4572334\\
FAX: 0039-055-4572364\\ E-mail: lusanna@fi.infn.it}

\today

\begin{abstract}

In this updated review of canonical ADM tetrad gravity in a family of globally hyperbolic asymptotically Minkowskian space-times without super-translations I show which is the status of the art in the search of a canonical basis adapted to the first class Dirac constraints and of the Dirac observables  of general relativity (GR) describing the tidal degrees of freedom of the gravitational field. In these space-times the asymptotic ADM Poincar\'e group replaces the Poincar\'e group of particle physics, there is a York canonical basis diagonalizing the York-Lichnerowicz approach
and a post-Minkowskian linearization is possible with the associated description of gravitational waves in the family of non-harmonic 3-orthogonal Schwinger time gauges.

Moreover I show that every fixation of the inertial gauge variables (i.e. the choice of a non-inertial frame) of every generally covariant formulation of GR is equivalent to a set of conventions for the metrology of the space-time (like the GPS ones near the Earth): for instance the freedom in clock synchronization is described by the inertial gauge variable York time (the trace of the extrinsic curvature of the instantaneous 3-spaces). This inertial gauge freedom and the non-Euclidean nature of the instantaneous 3-spaces required by the equivalence principle are connected with the dark side of the universe and could explain the presence of dark matter or at least part of it by means of the adoption of suitable metrical conventions for the ICRS celestial reference system. Also some comments on a canonical
quantization of GR  coherent with this viewpoint are done.

\end{abstract}

\bigskip

\maketitle

\section{Introduction}

The understanding of gravity and of its quantization is one of the most important challenges of contemporary theoretical physics. At the classical level inside the Solar System Einstein GR remains a satisfactory description of all the aspects of gravity \cite{a1}. Instead at the astrophysical and cosmological levels the dominance of dark matter and dark energy with respect to ordinary matter has induced the introduction of many possible alternatives to Einstein GR, like for instance $f(R)$ theories (see for instance Refs.\cite{a2}). Most of these theories are generally covariant like Einstein GR.
\medskip

In this paper I make a review of recent developments in a well
defined approach to classical canonical tetrad gravity and of their
implications for relativistic metrology in astrophysics. This is an upgrading of the old reviews \cite{a3}, which
contained also a survey of classical relativistic mechanics in non-inertial frames of Minkowski space-time
in presence of an electro-magnetic field. For an independent
review on classical canonical gravity see the first chapter of Ref.\cite{a4}.

\medskip

In a series of papers \cite{a5,a6,a7,a8} I looked to the existing
Hamiltonian formulations of metric and tetrad gravity \footnote{Tetrad gravity is needed for the
description of fermion fields: its formulation was introduced for the first time in
Ref.\cite{a9}.} by taking into account all the aspects of Dirac theory of constraints \cite{a10,a11}.

\medskip

The use of Hamiltonian methods restricts the class of Einstein
space-times to the {\it globally hyperbolic} ones, in which there is
a global notion of a mathematical time parameter. The space-times
must also be {\it topologically trivial}. At this level there are
two classes of physically inequivalent space-times with a completely
different dynamical interpretation:
\medskip

A) {\it Spatially compact space-times without boundary} - In them
the canonical Hamiltonian is zero and the Dirac Hamiltonian is a
linear combination of first class constraints. This fact gives rise
to a {\it frozen picture} without a global evolution (the Dirac
Hamiltonian generates only Hamiltonian gauge transformations; in the
abstract reduced phase space, quotient with respect to such gauge
transformations, the reduced Hamiltonian is zero). This class of
space-times fits well with Machian ideas (no boundary conditions),
with interpretations in which there is no physical time (see for
instance Ref.\cite{a12}) and is used in loop quantum gravity.

\medskip

B) {\it Asymptotically flat space-times} - In them we have the
asymptotic symmetries of the SPI group \cite{a13}
(direction-dependent asymptotic Killing symmetries). If we restrict
this class of space-times to those {\it not containing
super-translations} \cite{a14}, the SPI group reduces to the {\it
asymptotic ADM Poincar\'e group} \footnote{For recent reviews on
this group see Refs.\cite{a15,a16,a17}.}: these space-times are
asymptotically Minkowskian \footnote{This class of space-times admits
ortho-normal tetrads and a spinor structure \cite{a18}.}  and in the limit of vanishing Newton
constant ($G = 0$) the ADM Poincar\'e group becomes the special
relativistic Poincar\'e group of the matter present in the
space-time (this is an important condition for the inclusion of
particle physics, whose properties are all connected with the
representations of this group in the inertial frames of Minkowski
space-time, into GR). In this restricted class the
canonical Hamiltonian is the ADM energy \cite{a5}, so that there is
no frozen picture (in the reduced phase space there is a non-zero
reduced Hamiltonian). In absence of matter a sub-class of these
space-times is the (singularity-free) family of
Chrstodoulou-Klainermann solutions of Einstein equations \cite{a19}
(they are near to Minkowski space-time in a norm sense and contain
gravitational waves).

\medskip

Since the equivalence principle forbids the existence of global
inertial frames, we had to define {\it global non-inertial frames}
(with instantaneous non-Euclidean 3-spaces with synchronized clocks)
and radar 4-coordinates \cite{a20} in this class of
space-times by adapting to GR the theory of  global
non-inertial frames in Minkowski space-time developed in Ref. \cite{a21,a22}.
\medskip

In Refs.\cite{a23,a24} there is a systematic study of canonical ADM
tetrad  gravity  (it derives from ADM gravity \cite{a25} if in its
Lagrangian the 4-metric is decomposed in terms of cotetrads)
in {\it globally hyperbolic, asymptotically Minkowskian space-times without
super-translations} with electrically charged positive-energy scalar
particles plus the electro-magnetic field as matter.\medskip

In this formulation   all the 14 constraints are first
class \footnote{I consider only space-times without Killing symmetries, because otherwise there would be extra (either first or second class) constraints, Hamiltonian counterpart of the Killing equations. Also the eigenvalues of the 3-metric of the instantaneous 3-spaces must be different: again the degenerate cases can be described by adding the equality of the eigenvalues as extra constraints.}
and I tried to find Shanmugadhasan canonical transformations \cite{a26,a27,a28} to new
canonical bases in which Abelianized forms of many constraints are
new momenta (canonical bases adapted to as many as possible
constraints). Due to the non-linearity of the super-Hamiltonian and
super-momentum constraints, their solution is not known and we are
still unable to find their Abelianization and a canonical basis adapted to
all the first class constraints of GR.\medskip

As a consequence I could only find a York canonical basis \cite{a23} adapted to 10 first class
constraints and diagonalizing the York-Lichnerowitz approach \cite{a29,a30,a31,a32}.
In this basis one can identify which are: a) the four quantities to be determined by the
super-Hamiltonian and super-momentum constraints; b) the gauge variables of GR describing inertial effects
(one of them is the York time, i.e. the trace of the extrinsic curvature of the instantaneous 3-spaces);
c) the two pairs of dynamical variables describing the tidal effects of GR (the two polarizations of gravitational waves in the linearized theory). Do to the use of radar 4-coordinates {\it all these quantities are
4-scalars with respect to the ordinary world 4-coordinates}.\medskip

Instead the Dirac observables (DO) (gauge invariant
under the Hamiltonian gauge transformations generated by all the first
class constraints) of the gravitational field are not known: we have
only statements about their existence \cite{a33,a34,a35}. They would be the two pairs of tidal
variables in a Shanmugadhasan canonical basis adapted to all the 14 first class constraints.
In our approach they would be 4-scalars like the observables  of the Hilbert-Einstein action,
whose gauge group are the 4-diffeomorphisms of the space-time. See Ref.\cite{a36,a37,a38} for what is
known on the connection between 4-diffeomorphisms and the Hamiltonian gauge group: only on
the space of solutions of Einstein equations there is an overlap of
the two notions of observable.
\medskip

Moreover it is possible \cite{a39} to find the Hamiltonian expression of the Riemann and Weyl curvature tensors
and to give the Hamiltonian formulation of the Newman-Penrose formalism \cite{a40}.\medskip

An extremely important (till now unnoticed) point is that the fixation of the gauge freedom of GR (and
of every generally covariant theory of gravity), i.e. the choice of the non-inertial frame and of its 4-coordinates, is nothing else that {\it the establishment of conventions
for relativistic metrology}, an operation done from atomic physicists, NASA engineers and astronomers
\cite{a41,42,a43,a44}. As shown in the third paper of Refs.\cite{a24} the inertial gauge freedom in the York time is connected with the existence of dark matter at the cosmological level \cite{a45}. It is possible that at least part of dark matter is a relativistic inertial effect eliminable with a suitable convention for the ICRS celestial reference system. Moreover the York time is also connected with dark energy.\medskip

After these cosmological considerations I will end with some comments on the quantization of canonical gravity:
it is argued that only the final 4-scalar DO's describing the tidal effects should be quantized but not the inertial gauge variables describing the freedom in non-inertial frames.

\bigskip

In Section II I discuss non-inertial frames and radar 4-coordinates
in asymptotically Minkowskian space-times.
In Section III I define canonical ADM tetrad gravity and its York
canonical basis is introduced in Section IV, where there is also a
description of the Hamilton equations and of the inertial gauge
variable York time. In Section V I introduce the family of
non-harmonic 3-orthogonal Schwinger time gauges, I define a
Hamiltonian Post-Minkowskian (HPM) linearization and I describe the
resulting gravitational waves (GW).
In Section VI I study the HPM
equations of motion for particles and their Post-Newtonian (PN)
expansion, showing that the inertial gauge variable York time allows to
interpret (at least part of) astrophysical dark matter as a relativistic inertial
effect.
In Section VII I discuss relativistic metrology, whose conventions are gauge fixings for the inertial gauge variables of every generally covariant description of GR identifying global non-inertial frames and their 4-coordinates in the
space-time.

Then there are some Conclusions containing a sketch of the
open problems to be investigated with the described approach.

\section{Non-Inertial Frames in Asymptotically Minkowskian Space-Times
and the Gravitational Field}

Let us consider globally hyperbolic, topologically
trivial, asymptotically Minkowskian space-times without
super-translations. In them one can define global non-inertial
frames with the same methodology introduced in special relativity (SR) \cite{a21}.
\medskip

Assume that the world-line $x^{\mu}(\tau)$ of an arbitrary time-like
observer \footnote{An observer, or better a mathematical observer, is an idealization of a
measuring apparatus containing an atomic clock and defining, by means of
gyroscopes, a set of spatial axes (and then a, maybe orthonormal, tetrad
with a convention for its transport) in each point of the world-line.}
carrying a standard atomic clock is given: $\tau$ is an arbitrary
monotonically increasing function of the proper time of this clock. Then one
gives an admissible 3+1 splitting of the asymptotically flat space-time, namely a nice
foliation with space-like instantaneous 3-spaces $\Sigma_{\tau}$. It is the
mathematical idealization of a protocol for clock synchronization: all the
clocks in the points of $\Sigma_{\tau}$ sign the same time of the atomic
clock of the observer \footnote{It is the non-factual idealization
(generalizing the existing protocols for building coordinate systems inside the future light-cone of a time-like observer) required by the Cauchy problem:
without it we cannot use the existence and unicity theorem for the solutions of partial differential equations
to predict the future}. The observer and the
foliation define a global non-inertial reference frame after a choice of
4-coordinates. On each 3-space $\Sigma_{\tau}$  one chooses curvilinear
3-coordinates $\sigma^r$ having the observer as origin.\medskip

The quantities $\sigma^A = (\tau; \sigma^r)$ are the 4-scalar and observer-dependent
\textit{radar 4-coordinates}, first introduced by Bondi \cite{a20}. \medskip

If $x^{\mu} \mapsto \sigma^A(x)$ is the coordinate transformation from
world 4-coordinates $x^{\mu}$ having the observer as origin to radar 4-coordinates, its inverse $\sigma^A
\mapsto x^{\mu} = z^{\mu}(\tau ,\sigma^r)$ defines the \textit{embedding}
functions $z^{\mu}(\tau ,\sigma^r)$ describing the 3-spaces $\Sigma_{\tau}$
as embedded 3-manifolds into the asymptotically flat space-time.
Let $z^{\mu}_A(\tau, \sigma^u) = \partial\, z^{\mu}(\tau, \sigma^u) / \partial\, \sigma^A$ denote the gradients of the
embedding functions with respect to the radar 4-coordinates.
The space-like 4-vectors $z^{\mu}_r(\tau ,\sigma^u)$ are tangent to $\Sigma_{\tau}$, so that the unit
time-like normal $l^{\mu}(\tau ,\sigma^u)$ is proportional to $\epsilon^{\mu}{}_{%
\alpha \beta\gamma}\, [z^{\alpha}_1\, z^{\beta}_2\, z^{\gamma}_3](\tau
,\sigma^u)$ ($\epsilon_{\mu\alpha\beta\gamma}$ is the Levi-Civita tensor). Instead $z^{\mu}_{\tau}(\tau, \sigma^u)$ is a time-like 4-vector skew with respect to the 3-spaces leaves of the foliation
\footnote{In SR, see Refs. \cite{a21,a22},
one has $z^{\mu}_{\tau}(\tau ,\sigma^r) = [N\, l^{\mu} + N^r\,
z^{\mu}_r](\tau ,\sigma^r)$ with $N(\tau ,\sigma^r) = \epsilon\,
[z^{\mu}_{\tau}\, l_{\mu}](\tau ,\sigma^r) = 1 + n(\tau, \sigma^r) > 0$ and $
N_r(\tau ,\sigma^r) = - \epsilon\, [z^{\mu}_{\tau}\, \eta_{\mu\nu}\, z_r^{\mu}](\tau ,\sigma^r)$ being the
lapse and shift functions respectively of the global non-inertial frame of Minkowski space-time so defined.
This decomposition holds also in GR.}.\medskip

In GR the dynamical fields are the components ${}^4g_{\mu\nu}(x)$ of
the 4-metric \footnote{Let us remark that the ten dynamical fields ${}^4g_{\mu\nu}(x)$ are
not only a (pre)potential for the gravitational field (like the
electro-magnetic and Yang-Mills fields are the potentials for
electro-magnetic and non-Abelian forces) but also determines the
{\it chrono-geometrical structure of space-time} through the line
element $ds^2 = {}^4g_{\mu\nu}\, dx^{\mu}\, dx^{\nu}$. Therefore the
4-metric teaches relativistic causality to the other fields: it says
to massless particles like photons and gluons which are the allowed
world-lines in each point of space-time. This basic property is lost
in every quantum field theory approach to gravity with a fixed
background 4-metric.} and not the  embeddings $x^{\mu} = z^{\mu}(\tau,
\sigma^r)$ defining the admissible 3+1 splittings of space-time like
in  the parametrized Minkowski theories of SR \cite{a3,a21,a22}. Now the gradients
$z^{\mu}_A(\tau, \sigma^r)$ of the embeddings give the transition
coefficients from radar to world 4-coordinates, so that the
components ${}^4g_{AB}(\tau, \sigma^r) = z^{\mu}_A(\tau, \sigma^r)\,
z^{\nu}_B(\tau, \sigma^r)\, {}^4g_{\mu\nu}(z(\tau, \sigma^r))$ of
the 4-metric will be the dynamical fields in the ADM action \cite{a25}.
Let us remark that {\it the ten quantities ${}^4g_{AB}(\tau, \sigma^r)$ are
4-scalars of the space-time due to the use of the 4-scalar radar 4-coordinates}.
The same happens for {\it all the components of  "radar tensors" (i.e. tensors expressed in radar 4-coordinates):
they are 4-scalars of the space-time}.
\medskip

The 4-metric ${}^4g_{AB}$ has signature $\sgn\, (+---)$ with $\sgn =
\pm$ (the particle physics, $\sgn = +$, and general relativity,
$\sgn = -$, conventions). Flat indices $(\alpha )$, $\alpha = o, a$ ($a = 1,2,3$),
are raised and lowered by the flat Minkowski metric
${}^4\eta_{(\alpha )(\beta )} = \sgn\, (+---)$. We define
${}^4\eta_{(a)(b)} = - \sgn\, \delta_{(a)(b)}$ with a
positive-definite Euclidean 3-metric $\delta_{(a)(b)}$. From now on we shall denote
the curvilinear 3-coordinates $\sigma^r$ with the notation $\vec
\sigma$ for the sake of simplicity. Usually the convention of sum
over repeated indices is used, except when there are too many
summations. The symbol $\approx$ means Dirac weak equality, while the symbol
$\cir$ means evaluated by using the equations of motion.

\bigskip

We shall work with the radar tetrads ${}^4E^A_{(\alpha)}(\tau, \vec
\sigma)$ and the radar cotetrads ${}^4E_A^{(\alpha)}(\tau, \vec \sigma)$. The original tetrads
are ${}^4E^{\mu}_{(\alpha)}(\tau, \vec \sigma) = z^{\mu}_A(\tau, \vec
\sigma)\, {}^4E^A_{(\alpha)}(\tau, \vec \sigma)$.

\medskip

Since the world-line of the time-like observer can be chosen as the
origin of a set of the spatial world coordinates, i.e.
$x^{\mu}(\tau) = (x^o(\tau); 0)$, it turns out that with this choice
the space-like surfaces of constant coordinate time $x^o(\tau) =
const.$ coincide with the dynamical instantaneous 3-spaces
$\Sigma_{\tau}$ with $\tau = const.$. By using asymptotic flat
tetrads $\epsilon^{\mu}_A = \delta^{\mu}_o\, \delta^{\tau}_A +
\delta^{\mu}_i\, \delta^i_A$ (with $\epsilon^A_{\mu}$ denoting the
inverse flat cotetrads) and by choosing a coordinate world time
$x^o(\tau) = x^o_o + \epsilon^o_{\tau}\, \tau = x^o_o + \tau$, one
gets the following preferred embedding corresponding to these given
world 4-coordinates $x^{\mu} = z^{\mu}(\tau, \vec \sigma) =
x^{\mu}(\tau) + \epsilon^{\mu}_r\, \sigma^r = \delta^{\mu}_o\, x^o_o
+ \epsilon^{\mu}_A\, \sigma^A$. This choice implies $z^{\mu}_A(\tau,
\vec \sigma) = \epsilon^{\mu}_A$ and ${}^4g_{\mu\nu}(x = z(\tau,
\vec \sigma)) = \epsilon^A_{\mu}\, \epsilon_{\nu}^B\, {}^4g_{AB}(\tau,
\vec \sigma)$.

\bigskip

As shown in Ref.\cite{a46}, the dynamical nature of space-time
implies that each solution (i.e. an Einstein 4-geometry) of
Einstein's equations (or of the associated ADM Hamilton equations)
dynamically selects a preferred family of 3+1 splitting of the
space-time, namely in GR the instantaneous 3-spaces (and therefore
the associated clock synchronization convention) are dynamically
determined modulo only one inertial gauge function. As we will show,
in the York canonical basis this function is the {\it York time},
namely the trace of the extrinsic curvature of the 3-space. While
in SR the gauge freedom in clock synchronization depends on four
basic gauge functions, the embeddings $z^{\mu}(\tau, \sigma^r)$, and
both the 4-metric and the whole extrinsic curvature tensor are
derived inertial potentials, in GR the extrinsic curvature
tensor of the 3-spaces is a mixture of dynamical (tidal) pieces and
inertial gauge variables playing the role of inertial potentials
(but only the York time is a freedom in the choice of the shape of
the 3-spaces as 3-sub-manifolds of the space-time).

\section{Canonical ADM Tetrad Gravity}

To define the canonical formalism the Einstein-Hilbert action for metric gravity must be replaced
with the ADM action (the two actions differ for a surface tern at
spatial infinity). In the chosen class of space-times the ten strong
ADM Poincare'  generators $P^A_{ADM}$, $J^{AB}_{ADM}$ (they are
fluxes through a 2-surface at spatial infinity) are given as
boundary conditions at spatial infinity.
As shown in  Ref.\cite{a5}, the Legendre
transform and the definition of a consistent canonical Hamiltonian
require the introduction of the DeWitt surface term at spatial
infinity: the final canonical Hamiltonian turns out to be the {\it
strong} ADM energy (a flux through a 2-surface at spatial infinity),
which is equal to the {\it weak} ADM energy (expressed as a volume
integral over the 3-space) plus constraints. Therefore there is not
a frozen picture but an evolution generated by a Dirac Hamiltonian
equal to the weak ADM energy plus a linear combination of the first
class constraints. Also the other strong ADM Poincar\'e generators
are replaced by their weakly equivalent weak form ${\hat
P}^A_{ADM}$, ${\hat J}^{AB}_{ADM}$.\medskip

In Ref.\cite{a5} it is also shown that the
boundary conditions on the 4-metric required by the absence of
super-translations imply that the only admissible 3+1 splittings of
space-time (i.e. the allowed global non-inertial frames) are the
{\it non-inertial rest frames}:  their 3-spaces are asymptotically
orthogonal to the weak ADM 4-momentum. Therefore we get ${\hat
P}^r_{ADM} \approx 0$ as the rest-frame condition of the 3-universe
with a mass and a rest spin fixed by the boundary conditions. Like
in SR the 3-universe can be visualized as a decoupled non-covariant
(non-measurable) external relativistic center of mass plus an
internal non-inertial rest-frame 3-space containing only relative
variables (see the first paper in Ref.\cite{a24}).

\bigskip

In tetrad gravity  the 4-metric is decomposed in terms of cotetrads,
${}^4g_{AB} = E_A^{(\alpha)}\, {}^4\eta_{(\alpha)(\beta)}\,
E^{(\beta)}_B$ and the ADM action, now a
functional of the 16 fields $E^{(\alpha)}_A(\tau, \sigma^r)$, is
taken as the action for ADM tetrad gravity. The
diffeonorphism group (the gauge group of GR) is enlarged with the O(3,1) gauge group of the
Newman-Penrose approach \cite{a40} (the extra gauge freedom acting
on the tetrads in the tangent space of each point of space-time and
reducing from 16 to 10 the number of independent fields like in
metric gravity). This leads to an interpretation of gravity based on
a congruence of time-like observers endowed with ortho-normal
tetrads: in each point of space-time the time-like axis is the  unit
4-velocity of the observer, while the spatial axes are a (gauge)
convention for observer's gyroscopes. This framework was developed
in Refs.\cite{a6,a7}.

\medskip

In this framework the configuration variables are cotetrads, which
are connected to cotetrads adapted to the 3+1 splitting of
space-time (so that the adapted time-like tetrad is the unit normal
to the 3-space $\Sigma_{\tau}$) by standard Wigner boosts for
time-like vectors \footnote{In each tangent plane to a point of
$\Sigma_{\tau}$ the point-dependent standard Wigner boost for
time-like Poincare' orbits $L^{(\alpha )}{}_{(\beta )}(V(z(\sigma
));\,\, {\buildrel \circ \over V}) = \delta^{(\alpha )}_{(\beta )} +
2 \sgn\, V^{(\alpha )}(z(\sigma ))\, {\buildrel \circ \over
V}_{(\beta )} - \sgn\, {{(V^{(\alpha )}(z(\sigma )) + {\buildrel
\circ \over V}^{(\alpha )})\, (V_{(\beta )}(z(\sigma )) + {\buildrel
\circ \over V}_{(\beta )})}\over {1 + V^{(o)}(z(\sigma ))}}\,
{\buildrel {def}\over =}\, L^{(\alpha )}{}_{(\beta
)}(\varphi_{(a)})$ sends the unit future-pointing time-like vector
${\buildrel o\over V}^{(\alpha )} = (1; 0)$ into the unit time-like
vector $V^{(\alpha )} = {}^4E^{(\alpha )}_A\, l^A = \Big(\sqrt{1 +
\sum_a\, \varphi^2_{(a)}}; \varphi^{(a)} = - \sgn\,
\varphi_{(a)}\Big)$, where $l^A$ is the unit future-pointing normal
to $\Sigma_{\tau}$. We have $L^{-1}(\varphi_{(a)}) = {}^4\eta\,
L^T(\varphi_{(a)})\, {}^4\eta = L(- \varphi_{(a)})$. As a
consequence, the flat indices $(a)$ of the adapted tetrads and
cotetrads and of the triads and cotriads on $\Sigma_{\tau}$
transform as Wigner spin-1 indices under point-dependent SO(3)
Wigner rotations $R_{(a)(b)}(V(z(\sigma ));\,\, \Lambda (z(\sigma
))\, )$ associated with Lorentz transformations $\Lambda^{(\alpha
)}{}_{(\beta )}(z)$ in the tangent plane to the space-time in the
given point of $\Sigma_{\tau}$. Instead the index $(o)$ of the
adapted tetrads and cotetrads is a local Lorentz scalar index.} of
parameters $\varphi_{(a)}(\tau, \sigma^r)$:
${}^4E_A^{(\alpha)} = L^{(\alpha)}{}_{(\beta)}( \varphi_{(a)})\,
{}^4{\buildrel o\over E}_A^{(\beta)}$.
The adapted tetrads and cotetrads   have the expression

\bea
 \eo^A_{(o)} &=& {1\over {1 + n}}\, (1; - \sum_a\, n_{(a)}\,
 {}^3e^r_{(a)}) = l^A,\qquad \eo^A_{(a)} = (0; {}^3e^r_{(a)}), \nonumber \\
 &&{}\nonumber  \\
 \eo^{(o)}_A &=& (1 + n)\, (1; \vec 0) = \sgn\, l_A,\qquad \eo^{(a)}_A
= (n_{(a)}; {}^3e_{(a)r}),
 \label{3.1}
 \eea

\noindent where ${}^3e^r_{(a)}$ and ${}^3e_{(a)r}$ are triads and
cotriads on $\Sigma_{\tau}$ and $n_{(a)} = n_r\, {}^3e^r_{(a)} =
n^r\, {}^3e_{(a)r}$ \footnote{Since we use the positive-definite
3-metric $\delta_{(a)(b)} $, we shall use only lower flat spatial
indices. Therefore for the cotriads we use the notation
${}^3e^{(a)}_r\,\, {\buildrel {def}\over =}\, {}^3e_{(a)r}$ with
$\delta_{(a)(b)} = {}^3e^r_{(a)}\, {}^3e_{(b)r}$.} are adapted shift
functions. In Eqs.(\ref{3.1}) $N(\tau, \vec \sigma) = 1 + n(\tau,
\vec \sigma) > 0$, with $n(\tau ,\vec \sigma)$ vanishing at spatial
infinity (absence of super-translations), so that $N(\tau, \vec
\sigma)\, d\tau$ is positive from $\Sigma_{\tau}$ to $\Sigma_{\tau +
d\tau}$, is the lapse function; $N^r(\tau, \vec \sigma) = n^r(\tau,
\vec \sigma)$, vanishing at spatial infinity (absence of
super-translations), are the shift functions.

\bigskip

The adapted tetrads $\eo^A_{(a)}$ are defined modulo SO(3) rotations
$\eo^A_{(a)} = \sum_b\, R_{(a)(b)}(\alpha_{(e)})\, {}^4{\buildrel
\circ \over {\bar E}}^A_{(b)}$, ${}^3e^r_{(a)} = \sum_b\,
R_{(a)(b)}(\alpha_{(e)})\, {}^3{\bar e}^r_{(b)}$, where
$\alpha_{(a)}(\tau ,\vec \sigma )$ are three point-dependent Euler
angles. After having chosen an arbitrary point-dependent origin
$\alpha_{(a)}(\tau ,\vec \sigma ) = 0$, we arrive at the following
adapted tetrads and cotetrads [${\bar n}_{(a)} = \sum_b\, n_{(b)}\,
R_{(b)(a)}(\alpha_{(e)})\,$, $\sum_a\, n_{(a)}\, {}^3e^r_{(a)} =
\sum_a\, {\bar n}_{(a)}\,
 {}^3{\bar e}^r_{(a)}$]

\bea
 {}^4{\buildrel \circ \over {\bar E}}^A_{(o)}
 &=& \eo^A_{(o)} = {1\over {1 + n}}\, (1; - \sum_a\, {\bar n}_{(a)}\,
 {}^3{\bar e}^r_{(a)}) = l^A,\qquad {}^4{\buildrel \circ \over
 {\bar E}}^A_{(a)} = (0; {}^3{\bar e}^r_{(a)}), \nonumber \\
 &&{}\nonumber  \\
 {}^4{\buildrel \circ \over {\bar E}}^{(o)}_A
 &=& \eo^{(o)}_A = (1 + n)\, (1; \vec 0) = \sgn\, l_A,\qquad
 {}^4{\buildrel \circ \over {\bar E}}^{(a)}_A
= ({\bar n}_{(a)}; {}^3{\bar e}_{(a)r}),
 \label{3.2}
 \eea

\noindent which we shall use as a reference standard.\medskip

The expressions for the general tetrad and for the 4-metric
\footnote{The 3-metric ${}^3g_{rs}$ has signature $(+++)$, so that we may put
all the flat 3-indices {\it down}. We have ${}^3g^{ru}\, {}^3g_{us}
= \delta^r_s$.} are

\bea
 {}^4E^A_{(\alpha )} &=& \eo^A_{(\beta )}\, L^{(\beta )}{}_{(\alpha
 )}(\varphi_{(a)}) = {}^4{\buildrel \circ \over {\bar E}}^A_{(o)}\,
 L^{(o)}{}_{(\alpha )}(\varphi_{(c)}) + \sum_{ab}\, {}^4{\buildrel \circ \over
 {\bar E}}^A_{(b)}\, R^T_{(b)(a)}(\alpha_{(c)})\,
 L^{(a)}{}_{(\alpha )}(\varphi_{(c)}),\nonumber \\
  {}^4g_{AB} &=& \eo^{(\alpha)}_A\, {}^4\eta_{(\alpha )(\beta )}\, \eo^{(\beta )}_B =
  {}^4{\buildrel \circ \over {\bar E}}^{(\alpha)}_A\,
 {}^4\eta_{(\alpha)(\beta)}\, {}^4{\buildrel \circ \over {\bar E}}^{(\beta)}_B,\nonumber \\
  {}^4g_{\tau\tau} &=& \sgn\, [(1 + n)^2 - {}^3g^{rs}\, n_r\,
 n_s] = \sgn\, [(1 + n)^2 - \sum_a\, {\bar n}^2_{(a)}],\qquad
 {}^4g_{\tau r} = - \sgn\, n_r = -\sgn\, \sum_a\, {\bar n}_{(a)}\,
 {}^3{\bar e}_{(a)r},\nonumber \\
  {}^4g_{rs} &=& -\sgn\, {}^3g_{rs} = - \sgn\, \sum_a\, {}^3e_{(a)r}\, {}^3e_{(a)s}
  = - \sgn\, \sum_a\, {}^3{\bar e}_{(a)r}\, {}^3{\bar e}_{(a)s},\nonumber \\
   \sqrt{- g } &=& \sqrt{|{}^4g|} = {{\sqrt{{}^3g}}\over {\sqrt{\sgn\,
 {}^4g^{\tau\tau}}}} = \sqrt{\gamma}\, (1 + n) = {}^3e\, (1 +
 n),\qquad {}^3g = \gamma =
 ({}^3e)^2,\quad {}^3e = det\, {}^3e_{(a)r}.\nonumber \\
 &&{}
  \label{3.3}
 \eea

\bigskip

The future-oriented unit normal to $\Sigma_{\tau}$ and the projector
on $\Sigma_{\tau}$ are $l_A = \sgn\, (1 + n)\, \Big(1;\, 0\Big)$,
${}^4g^{AB}\, l_A\, l_B = \sgn $, $l^A = \sgn\, (1 + n)\,
{}^4g^{A\tau} = {1\over {1 + n}}\, \Big(1;\, - n^r\Big) = {1\over {1
+ n}}\, \Big(1;\, - \sum_a\, {\bar n}_{(a)}\, {}^3{\bar
e}_{(a)}^r\Big)$, ${}^3h^B_A = \delta^B_A - \sgn\, l_A\, l^B$.

\bigskip

Each 3+1 splitting of (either Minkowski or asymptotically
Minkowskian) space-time, i.e. each global non-inertial frame, has
two associated congruences of time-like observers:\hfill\break

i) The congruence of the Eulerian observers with the unit normal
$l^{\mu}(\tau, \vec \sigma) = \Big(z^{\mu}_A\, l^A\Big)(\tau, \vec
\sigma)$ to the 3-spaces as unit 4-velocity. The world-lines of
these observers are the integral curves of the unit normal and in
general are not geodesics. In adapted radar 4-coordinates  the
Eulerian observers carry the contro-variant ($l^A(\tau, \vec
\sigma)$, ${}^4{\buildrel \circ \over {\bar E}}^A_{(a)}(\tau, \vec
\sigma)$) and covariant ($l_A(\tau, \vec \sigma)$, ${}^4{\buildrel
\circ \over {\bar E}}_{(a)A}(\tau, \vec \sigma)$) orthonormal
tetrads defined in of Eqs.(\ref{3.2}).

\bigskip

ii) The skew congruence with unit 4-velocity $v^{\mu}(\tau, \vec
\sigma) = \Big(z^{\mu}_A\, v^A\Big)(\tau, \vec \sigma)$ (in general
it is not surface-forming, i.e. it has a non-vanishing vorticity).
The observers of the skew congruence have the world-lines (integral
curves of the 4-velocity) defined by $\sigma^r = const.$ for every
$\tau$, because the unit 4-velocity tangent to the flux lines
$x^{\mu}_{{\vec \sigma}_o}(\tau) = z^{\mu}(\tau, {\vec \sigma}_o)$
is $v^{\mu}_{{\vec \sigma}_o}(\tau) = z^{\mu}_{\tau}(\tau, {\vec
\sigma}_o)/\sqrt{\sgn\, {}^4g_{\tau\tau}(\tau, {\vec \sigma}_o)}$.
They carry the adapted contro-variant and covariant orthonormal
tetrads (${\cal V}^A_{(a)}(\tau, \vec \sigma)$ are not tangent to
the 3-spaces $\Sigma_{\tau}$ like ${}^4{\buildrel \circ \over {\bar
E}}^A_{(a)}(\tau, \vec \sigma)$ of Eqs.(\ref{3.2}))

\bea
 v^A(\tau, \vec \sigma) &=& {{(1; 0)}\over {\sqrt{(1 + n)^2 - \sum_a\,
 {\bar n}^2_{(a)}}}}(\tau, \vec \sigma),\nonumber \\
 {\cal V}^A_{(a)}(\tau, \vec \sigma) &=&  \Big({{{\bar
 n}_{(a)}}\over {(1 + n)^2}}; \sum_b\, (\delta_{(a)(b)} - {{{\bar n}_{(a)}\,
 {\bar n}_{(b)}}\over {(1 + n)^2}})\, {}^3{\bar e}^r_{(b)}\Big)(\tau,
 \vec \sigma),\nonumber \\
 &&{}\nonumber \\
  \sgn\, v_A(\tau, \vec \sigma) &=&  \Big(\sqrt{(1 + n)^2 - \sum_c\,
 {\bar n}^2_{(c)}}; {{- {\bar n}_{(a)}\, {}^3{\bar e}_{(a)r}}\over
 {\sqrt{(1 + n)^2 - \sum_c\, {\bar n}^2_{(c)}}}}\Big)(\tau, \vec
 \sigma),\nonumber \\
  {\cal V}_{(a)A}(\tau, \vec \sigma) &=& \Big(0; {}^3{\bar
 e}_{(a)r}\Big)(\tau, \vec \sigma).
 \label{3.4}
 \eea

\bigskip

The 16 configurational variables in the ADM action are
$\varphi_{(a)}$, $1 + n$, $n_{(a)}$, ${}^3e_{(a)r}$. There are ten
primary constraints (the vanishing of the 7 momenta of boosts, lapse
and shift variables plus three constraints describing the rotation
on the flat indices $(a)$ of the cotriads) and four secondary ones
(the super-Hamiltonian and super-momentum constraints): all of them
are first class in the phase space spanned by 16+16 fields. This
implies that there are 14 gauge variables describing {\it inertial
effects} and 2 canonical pairs of physical degrees of freedom
describing the {\it tidal effects} of the gravitational field
(namely gravitational waves in the weak field limit). In this
canonical basis only the momenta ${}^3\pi^r_{(a)}$ conjugated to the
cotriads are not vanishing. The basis of canonical variables for this formulation of tetrad
gravity, naturally adapted to 7 of the 14 first-class constraints,
is

\beq
 \begin{minipage}[t]{3cm}
\begin{tabular}{|l|l|l|l|} \hline
$\varphi_{(a)}$ & $n$ & $\bar n_{(a)}$ & ${}^3e_{(a)r}$ \\ \hline $
\pi_{\varphi_{(a)}}\, \approx 0$ & $\pi_n\, \approx 0$ &
$\pi_{n_{(a)}}\, \approx 0 $ & ${}^3{ \pi}^r_{(a)}$
\\ \hline
\end{tabular}
\end{minipage}
 \label{3.5}
 \eeq

In the next Section I will show a more convenient canonical basis in which the
configuration variables are $\alpha_{(a)}$, $\varphi_{(a)}$, $1 + n$, ${\bar n}_{(a)}$, ${}^3{\bar e}_{(a)r}$.

\medskip

From Eqs.(5.5) of  Ref.\cite{a7} we assume the
following (direction-independent, so to kill super-translations)
boundary conditions at spatial infinity ($r = \sqrt{\sum_r\,
(\sigma^r)^2}$; $\epsilon > 0$; $M = const.$): $n(\tau, \sigma^r)
\rightarrow_{r\, \rightarrow\, \infty}\, O(r^{- (2 + \epsilon )})$,
$\pi_n(\tau, \sigma^r) \rightarrow_{r\, \rightarrow\, \infty}\,
O(r^{-3})$, $n_{(a)}(\tau, \sigma^r) \rightarrow_{r\, \rightarrow\,
\infty}\, O(r^{- \epsilon })$, $\pi_{n_{(a)}}(\tau, \sigma^r)
\rightarrow_{r\, \rightarrow\, \infty}\, O(r^{-3})$,
$\varphi_{(a)}(\tau, \sigma^r) \rightarrow_{r\, \rightarrow\,
\infty}\, O(r^{- (1 + \epsilon)})$, $\pi_{\varphi_{(a)}}(\tau,
\sigma^r) \rightarrow_{r\, \rightarrow\, \infty}\, O(r^{-2})$,
${}^3e_{(a)r}(\tau, \sigma^r) \rightarrow_{r\, \rightarrow\,
\infty}\, \Big(1 + {M\over {2 r}}\Big)\, \delta_{ar} + O(r^{-
3/2})$, ${}^3\pi^r_{(a)}(\tau, \sigma^r) \rightarrow_{r\,
\rightarrow\, \infty}\, O(r^{- 5/2})$.

\vfill\eject

\section{The York Canonical Basis}

In Ref.\cite{a23}  a canonical transformation to a
canonical basis adapted to ten of the first class constraints was found. It
implements the York map of Ref.\cite{a31} (in the cases in which the
3-metric ${}^3g_{rs}$ has three distinct eigenvalues) and
diagonalizes the York-Lichnerowicz approach. Its final form is ($\alpha_{(a)}(\tau, \sigma^r)$ are the
Euler angles of the previous Section;  $V_{ua} {\buildrel {def}\over =} \sum_v\, V_{uv}\, \delta_{v(a)}$)

\bea
 &&\begin{minipage}[t]{4 cm}
\begin{tabular}{|ll|ll|l|l|l|} \hline
$\varphi_{(a)}$ & $\alpha_{(a)}$ & $n$ & ${\bar n}_{(a)}$ &
$\theta^r$ & $\tilde \phi$ & $R_{\bar a}$\\ \hline
$\pi_{\varphi_{(a)}} \approx0$ &
 $\pi^{(\alpha)}_{(a)} \approx 0$ & $\pi_n \approx 0$ & $\pi_{{\bar n}_{(a)}} \approx 0$
& $\pi^{(\theta )}_r$ & $\pi_{\tilde \phi} = {{c^3}\over {12\pi G}}\, {}^3K$ & $\Pi_{\bar a}$ \\
\hline
\end{tabular}
\end{minipage}\nonumber \\
 &&{}\nonumber \\
 &&{}\nonumber \\
 &&{}^3e_{(a)r} = \sum_b\, R_{(a)(b)}(\alpha_{(c)})\, {}^3{\bar e}_{(b)r},\qquad
 {}^3{\bar e}_{(a)r} = V_{ra}(\theta^i)\,
 {\tilde \phi}^{1/3}\, e^{\sum_{\bar a}^{1,2}\, \gamma_{\bar aa}\, R_{\bar a}},\nonumber \\
 &&{}\nonumber \\
 &&{}^4g_{\tau\tau} = \sgn\, [(1 + n)^2 - \sum_a\, {\bar n}^2_{(a)}],
 \qquad {}^4g_{\tau r} = - \sgn\, \sum_a\, {\bar
 n}_{(a)}\, {}^3{\bar e}_{(a)r},\nonumber \\
 &&{}^4g_{rs} = - \sgn\, {}^3g_{rs} = - \sgn\, {\tilde \phi}^{2/3}\,
 \sum_a\, V_{ra}(\theta^i)\, V_{sa}(\theta^i)\, Q^2_a,\qquad
 Q_a = e^{ \sum_{\bar a}^{1,2}\, \gamma_{\bar aa}\, R_{\bar
 a}},\nonumber \\
 &&{}\nonumber \\
 \label{4.1}
 \eea

The set of numerical parameters $\gamma_{\bar aa}$ satisfies
\cite{a5} $\sum_u\, \gamma_{\bar au} = 0$, $\sum_u\, \gamma_{\bar a
u}\, \gamma_{\bar b u} = \delta_{\bar a\bar b}$, $\sum_{\bar a}\,
\gamma_{\bar au}\, \gamma_{\bar av} = \delta_{uv} - {1\over 3}$.
Each solution of these equations defines a different York canonical
basis.

\medskip

This canonical basis has been found  due to the fact that the
3-metric $ {}^3g_{rs}$ is a real symmetric $3 \times 3$ matrix,
which may be diagonalized with an {\it orthogonal} matrix
$V(\theta^r)$, $V^{-1} = V^T$ ($\sum_u\, V_{ua}\, V_{ub} =
\delta_{ab}$, $\sum_a\, V_{ua}\, V_{va} = \delta_{uv}$, $\sum_{uv}\,
\epsilon_{wuv}\, V_{ua}\, V_{vb} = \sum_c\, \epsilon_{abc}\,
V_{cw}$), $det\, V = 1$, depending on three parameters $\theta^r$. If we choose these three gauge
parameters to be Euler angles ${\hat \theta}^i(\tau, \vec \sigma)$,
we get a description of the 3-coordinate systems on $\Sigma_{\tau}$
from a local point of view, because they give the orientation of the
tangents to the three 3-coordinate lines through each point.
However, for the calculations (see Refs.\cite{a24}) it is more
convenient to choose the three gauge parameters as first kind
coordinates $\theta^i(\tau, \vec \sigma)$ ($- \infty < \theta^i < +
\infty$) on the O(3) group manifold, so that by definition we have
$V_{ru}(\theta^i) = \Big(e^{- \sum_i\, {\hat T}_i\,
\theta^i}\Big)_{ru}$, where $({\hat T}_i)_{ru} = \epsilon_{rui}$ are
the generators of the o(3) Lie algebra in the adjoint
representation, and the Euler angles may be expressed as ${\hat
\theta}^i = f^i(\theta^n)$. The Cartan matrix has the form
$A(\theta^n) = {{1 - e^{- \sum_i\, {\hat T}_i\, \theta^i} }\over
{\sum_i\, {\hat T}_i\, \theta^i}}$ and satisfies $A_{ri}(\theta^n)\,
\theta^i = \delta_{ri}\, \theta^i$; $B(\theta^i) =
A^{-1}(\theta^i)$.\medskip

From now on for the sake of notational simplicity we shall use $\vec
\sigma$ for the curvilinear coordinates $\sigma^r$ and $V$ for
$V(\theta^i)$.

\bigskip

This canonical transformation realizes a {\it York map} because the
gauge variable $\pi_{\tilde \phi}$ (describing the freedom in the
choice of the trace of the extrinsic curvature of the instantaneous
3-spaces $\Sigma_{\tau}$) is proportional to {\it York internal
extrinsic time} ${}^3K$. It is the only gauge variable among the
momenta: this is a reflex of the Lorentz signature of space-time,
because $\pi_{\tilde \phi}$ and $\theta^n$ can be used as a set of
4-coordinates \cite{a46}. The York time describes the effect of
gauge transformations producing a deformation of the shape of the
3-space along the 4-normal to the 3-space as a 3-sub-manifold of
space-time.

Its conjugate variable, to be determined by the super-Hamiltonian
constraint, is $\tilde \phi  = {}^3\bar e = \sqrt{det\,
{}^3g_{rs}}$, which is proportional to {\it Misner's internal
intrinsic time}; moreover $\tilde \phi$ is the {\it 3-volume
density} on $\Sigma_{\tau}$: $V_R = \int_R d^3\sigma\, \tilde \phi$,
$R \subset \Sigma_{\tau}$. Since we have ${}^3g_{rs} = {\tilde
\phi}^{2/3}\, {}^3{\hat g}_{rs}$ with $det\, {}^3{\hat g}_{rs} = 1$,
$\tilde \phi$ is also called the {\it conformal factor} of the
3-metric.

\medskip

The two pairs of canonical variables $R_{\bar a}$, $\Pi_{\bar a}$,
$\bar a = 1,2$, describe the generalized {\it tidal effects}, namely
the independent physical degrees of freedom of the gravitational
field. They are 3-scalars on $\Sigma_{\tau}$ and
the configuration tidal variables $R_{\bar a}$ depend {\it only on
the eigenvalues of the 3-metric}. They are DO's {\it
only} with respect to the gauge transformations generated by 10 of
the 14 first class constraints. Let us remark that, if we fix
completely the gauge and we go to Dirac brackets, then the only
surviving dynamical variables $R_{\bar a}$ and $\Pi_{\bar a}$ become
two pairs of {\it non canonical} Dirac observables for that gauge:
the two pairs of canonical Dirac observables have to be found as a
Darboux basis of the copy of the reduced phase space identified by
the gauge and they will be (in general non-local) functionals of the
$R_{\bar a}$, $\Pi_{\bar a}$ variables. \medskip

Since the variables $\tilde \phi$  and $\pi_i^{(\theta )}$
are determined by the super-Hamiltonian (i.e.
the Lichnerowicz equation) and super-momentum constraints respectively, the {\it
arbitrary gauge variables} are $\alpha_{(a)}$, $\varphi_{(a)}$,
$\theta^i$, $\pi_{\tilde \phi}$, $n$ and ${\bar n}_{(a)}$. As shown
in Refs.\cite{a23}, they describe the following generalized {\it
inertial effects}:\medskip

a) $\alpha_{(a)}(\tau ,\vec \sigma )$ and $\varphi_{(a)}(\tau ,\vec
\sigma )$ are the 6 configuration variables parametrizing the O(3,1)
gauge freedom in the choice of the tetrads in the tangent plane to
each point of $\Sigma_{\tau}$ and describe the arbitrariness in the
choice of a tetrad to be associated to a time-like observer, whose
world-line goes through the point $(\tau ,\vec \sigma )$. They fix
{\it the unit 4-velocity of the observer and the conventions for the
orientation of three gyroscopes and their transport along the
world-line of the observer}. The  {\it Schwinger time gauges} are
defined by the gauge fixings $\alpha_{(a)}(\tau, \vec \sigma)
\approx 0$, $\varphi_{(a)}(\tau, \vec \sigma) \approx 0$.
\medskip

b) $\theta^i(\tau ,\vec \sigma )$ (depending only on the 3-metric)
describe the arbitrariness in the choice of the 3-coordinates in the
instantaneous 3-spaces $\Sigma_{\tau}$ of the chosen non-inertial
frame  centered on an arbitrary time-like observer. Their choice
will induce a pattern of {\it relativistic inertial forces} for the
gravitational field, whose potentials are the functions
$V_{ra}(\theta^i)$ present in the weak ADM energy ${\hat E}_{ADM}$.
\medskip

c) ${\bar n}_{(a)}(\tau ,\vec \sigma )$, the shift functions,
describe which points on different instantaneous 3-spaces have the
same numerical value of the 3-coordinates. They are the inertial
potentials describing the effects of the non-vanishing off-diagonal
components ${}^4g_{\tau r}(\tau ,\vec \sigma )$ of the 4-metric,
namely they are the {\it gravito-magnetic potentials} \footnote{In
the post-Newtonian approximation in harmonic gauges they are the
counterpart of the electro-magnetic vector potentials describing
magnetic fields \cite{a32}: A) $N = 1 + n$, $n\, {\buildrel
{def}\over =}\, - {{4\, \sgn}\over {c^2}}\, \Phi_G$ with $\Phi_G$
the {\it gravito-electric potential}; B) $n_r\, {\buildrel
{def}\over =}\, {{2\, \sgn}\over {c^2}}\, A_{G\, r}$ with $A_{G\,
r}$ the {\it gravito-magnetic} potential; C) $E_{G\, r} =
\partial_r\, \Phi_G - \partial_{\tau}\, ({1\over 2}\, A_{G\, r})$ (the {\it
gravito-electric field}) and $B_{G\, r} = \epsilon_{ruv}\,
\partial_u\, A_{G\, v} = c\, \Omega_{G\, r}$ (the {\it
gravito-magnetic field}). Let us remark that in arbitrary gauges the
analogy with electro-magnetism  breaks down.} responsible of effects
like the dragging of inertial frames (Lens-Thirring effect)
in the post-Newtonian approximation. The shift functions
are determined by the $\tau$-preservation of the gauge fixings
determining the gauge variables $\theta^i(\tau, \vec \sigma)$.
\medskip

d) $\pi_{\tilde \phi}(\tau ,\vec \sigma )$, i.e. the York time
${}^3K(\tau ,\vec \sigma )$, describes the non-dynamical
arbitrariness in the choice of the convention for the
synchronization of distant clocks which remains in the transition
from SR to GR. Since the York time is present in the Dirac
Hamiltonian, it is a {\it new inertial potential} connected to the
problem of the relativistic freedom in the choice of the {\it shape
of the instantaneous 3-space}, which has no Newtonian analogue (in
Galilei space-time time is absolute and there is an absolute notion
of Euclidean 3-space). Its effects are completely unexplored.

\medskip

e) $1 + n(\tau ,\vec \sigma )$, the lapse function appearing in the
Dirac Hamiltonian, describes the arbitrariness in the choice of the
unit of proper time in each point of the simultaneity surfaces
$\Sigma_{\tau}$, namely how these surfaces are packed in the 3+1
splitting. The lapse function is determined by the
$\tau$-preservation of the gauge fixing for the gauge variable
${}^3K(\tau, \vec \sigma)$.

\bigskip

\bigskip

The extrinsic curvature tensor of the 3-space $\Sigma_{\tau}$ has
the expression

\bea
  {}^3K_{rs} &=&  - {{4\pi\, G}\over {c^3}}\, {\tilde \phi}^{-1/3}\,
 \Big(\sum_a\, Q^2_a\, V_{ra}\, V_{sa}\, [2\, \sum_{\bar b}\, \gamma_{\bar ba}\,
 \Pi_{\bar b} -  \tilde \phi\, \pi_{\tilde \phi}] +\nonumber \\
 &+& \sum_{ab}\, Q_a\, Q_b\, (V_{ra}\, V_{sb} +
 V_{rb}\, V_{sa})\, \sum_{twi}\, {{\epsilon_{abt}\,
 V_{wt}\, B_{iw}\, \pi_i^{(\theta )}}\over {
 Q_b\, Q^{-1}_a  - Q_a\, Q^{-1}_b}} \Big).
 \label{4.2}
 \eea

\medskip

As shown in Eqs.(2.11)-(2.16) of the first paper in
Refs.\cite{a24}, if we use radar 4-coordinates, the covariant unit
normal $\sgn\, l_A = (1 + n)\, (1; 0)$, i.e. the 4-velocity of the Eulerian observers,  has the following covariant
derivative

\beq
 {}^4\nabla_A\,\, \sgn\, l_B = \sgn\, l_A\, {}^3a_B + \sigma_{AB} +
 {1\over 3}\, \theta\, ({}^4g_{AB} - \sgn\, l_A\, l_B) - \omega_{AB}.
 \label{4.3}
 \eeq
\medskip

The quantities appearing in Eqs.(\ref{4.3}) are:\medskip

1) the {\it acceleration} ${}^3a_A$ of the Eulerian observers
(${}^3a_r = \partial_r\, ln\, (1 + n)$, ${}^3a_{\tau} =  {}^3a_r\,
{}^3{\bar e}^r_{(a)}\, {\bar n}_{(a)}$);\medskip

2)  their {\it expansion}, which coincides with the {\it York
external time} \footnote{In cosmology it is proportional
 to the {\it Hubble parameter} $H = {1\over 3}\, \theta$
and determines the dimensionless (cosmological) {\it deceleration
parameter} $q = - 3\, \theta^{-2}\, l^A\,
\partial_A\, \theta - 1$} : $\theta = {}^4\nabla_A\,\, l^A
 = - \sgn\, {}^3K =  - \sgn\, {{12\pi\, G}\over {c^3}}\, \pi_{\tilde \phi}$;\medskip

3) their {\it shear} $\sigma_{AB}$, whose components $\sigma_{(\alpha)(\beta)}$
along the tetrads (\ref{5.2}) turn out to be $\sigma_{(o)(o)} =
\sigma_{(o)(a)} = 0$ and $\sigma_{(a)(b)} = \sigma_{(b)(a)} =
({}^3K_{rs} - {1\over 3}\, {}^3g_{rs}\, {}^3K)\, {}^3{\bar
e}^r_{(a)}\, {}^3{\bar e}^s_{(b)}$ with $ \sum_a\, \sigma_{(a)(a)} =
0$. $\sigma_{(a)(b)}$ depends upon $\theta^r$, $\tilde \phi$,
$R_{\bar a}$, $\pi^{(\theta )}_r$ and $\Pi_{\bar a}$.

Instead the definition of Eulerian observers implies that their {\it
vorticity} or {\it twist} vanishes because the congruence is
surface-forming: $\omega_{AB} = - \omega_{BA} = 0$.

\bigskip

Then the following results can be obtained

\bea
 \tilde \phi\, \sigma_{(a)(a)} &=& - {{8\pi\, G}\over {c^3}}\,
 \sum_{\bar a}\, \gamma_{\bar aa}\, \Pi_{\bar a},\,
 \rightarrow\, \Pi_{\bar a} = - {{c^3}\over {8\pi\, G}}\, \tilde \phi\,
 \sum_a\, \gamma_{\bar aa}\, \sigma_{(a)(a)},\nonumber \\
 &&{}\nonumber \\
 \tilde \phi\, \sigma_{(a)(b)}{|}_{a \not= b} &=& - {{8\pi\, G}\over
 {c^3}}\, \sum_{tw}\, {{\epsilon_{abt}\, V_{wt}}\over
 {Q_b\, Q_a^{-1} - Q_a\, Q_b^{-1}}}\, \sum_i\, B_{iw}\,
 \pi_i^{(\theta )},\nonumber \\
 &&{}\nonumber \\
 \Rightarrow&& \pi_i^{(\theta )} =  {{c^3}\over {8\pi\, G}}\, \tilde
 \phi\, \sum_{wtab}\, A_{wi}\, V_{wt}\, Q_a\, Q_b^{-1}\, \epsilon_{tab}\,
 \sigma_{(a)(b)}{|}_{a\not= b},\nonumber \\
 &&{}\nonumber \\
 {}^3K_{rs} &=& - {{\sgn}\over 3}\, {}^3g_{rs}\, \theta +
 \sigma_{(a)(b)}\, {}^3{\bar e}_{(a)r}\, {}^3{\bar e}_{(b)s}.
 \label{4.4}
 \eea

\bigskip

Therefore the diagonal elements of the shear of the Eulerian
observers describe the tidal momenta $\Pi_{\bar a}$, while the
non-diagonal elements determine the variables $\pi_i^{(\theta )}$,
determined by the super-momentum constraints. Moreover their
expansion $\theta$  is the inertial gauge variable  determining the
non-dynamical part (general relativistic gauge freedom in clock
synchronization) of the shape of the instantaneous 3-spaces
$\Sigma_{\tau}$.

\bigskip

See the first paper in Refs.\cite{a24} for the expression of the
super-momentum constraints  ${\cal H}_{(a)}(\tau, \vec \sigma)
\approx 0$ [Eqs.(3.41)-(3.42)] and of the super-Hamiltonian
constraint ${\cal H}(\tau, \vec \sigma) \approx 0$ (the Lichnerowicz
equation) [Eqs.(3.44)-(3.45)]. The weak ADM energy is given in Eqs.
(3.43)-(3.45) of that paper (while the other weak Poincar\'e
generators are given in Eqs.(3.47)): in it there is a negative
kinetic term proportional to $({}^3K)^2$ (the York time is a
momentum!), vanishing only in the gauges ${}^3K(\tau, \vec \sigma) =
0$. It comes from the bilinear in momenta present both in the
super-Hamiltonian and in the weak ADM energy: it was known that this
quadratic form was not definite positive but only in the York
canonical basis this can be made explicit. The expression of the
weak ADM energy in terms of the expansion ($\theta = - \sgn\, {}^3K
= - \sgn\, {{12\pi G}\over {c^3}}\, \pi_{\tilde \phi}$) and shear of
the Eulerian observers is

 \bea
 {\hat E}_{ADM} &=& c\, \int d^3\sigma\, \Big[{\check {\cal M}} -
  {{c^3}\over {16\pi\, G}}\, {\cal S} + {{4\pi\, G}\over {c^3}}\,
  {\tilde \phi}^{-1}\, \sum_{\bar b}\, \Pi^2_{\bar b} +\nonumber \\
  &+& \tilde \phi\, \Big( {{c^3}\over {16\pi\, G}}\,
 \sum_{ab, a\not= b}\, \sigma^2_{(a)(b)}
 - {{6\pi\, G}\over {c^3}}\, \pi^2_{\tilde \phi}\Big)
 \,\, \Big](\tau ,\vec \sigma ),
 \label{4.5}
 \eea

\noindent where ${\check {\cal M}} = \tilde \phi\, (1 + n)^2\,
T^{\tau\tau}$ is the energy-mass density of the matter (with
energy-momentum tensor $T^{AB}$) and ${\cal S}(\tilde \phi,
\theta^i, R_{\bar a})$ is an inertial potential depending on the
choice of the 3-coordinates in the 3-space (it is the
$\Gamma-\Gamma$ term in the scalar 3-curvature of the 3-space).

Finally the Dirac Hamiltonian is

\bea
  H_D&=& {1\over c}\, {\hat E}_{ADM} + \int d^3\sigma\, \Big[ n\,
{\cal H} - {\bar n}_{(a)}\, {\cal H}_{(a)}\Big](\tau ,\vec \sigma )
+ \lambda_r(\tau )\, {\hat P}^r_{ADM} +\nonumber \\
 &+&\int d^3\sigma\, \Big[\lambda_n\, \pi_n + \lambda_{
{\bar n}_{(a)}}\, \pi_{{\bar n}_{(a)}} + \lambda_{\varphi_{(a)}}\,
\pi_{ \varphi_{(a)}} + \lambda_{\alpha_(a)}\,
\pi^{(\alpha)}_{(a)}\Big](\tau ,\vec \sigma ),
 \label{4.6}
 \eea

\noindent where the $\lambda_{...}(\tau, \vec \sigma)$'s are Dirac
multipliers. In particular the Dirac multiplier $\lambda_r(\tau)$
implements the rest frame condition ${\hat P}^r_{ADM} \approx 0$.

\medskip

In the York canonical basis, where the super-momentum and
super-Hamiltonian constraints are coupled {\it elliptic} equations
on the 3-space $\Sigma_{\tau}$, the Hamilton equations generated by
this Dirac Hamiltonian  (replacing the standard 12 ADM equations and
the matter equations ${}^4\nabla_A\, T^{AB} = 0$) are divided in
four groups:\medskip

A) the contracted Bianchi identities, namely the evolution equations
for the solutions $\tilde \phi$ and $\pi_i^{(\theta)}$ of the
constraints (they say that given a solution of the constraints on a
Cauchy surface, it remains a solution also at later times);\medskip

B) the evolution equation for the four basic gauge variables
$\theta^i$ and ${}^3K$: these equations determine the lapse and the
shift functions once four gauge-fixings for the basic gauge
variables are given;\medskip

C) the {\it hyperbolic} evolution equations for the tidal variables
$R_{\bar a}$, $\Pi_{\bar a}$;\medskip

D) the Hamilton equations for matter, when present.\medskip

Once a gauge is completely fixed by giving the six gauge-fixings for
the O(3,1) variables $\varphi_{(a)}$, $\alpha_{(a)}$ (choice of the
tetrads and of their transport) and four gauge-fixings for
$\theta^i$ (choice of the 3-coordinates on the 3-space) and ${}^3K$
(determination of the shape of the 3-space as a 3-sub-manifold of
space-time by means of a clock synchronization convention), the
Hamilton equations become a deterministic set of coupled PDE's for
the lapse and shift functions (secondary inertial gauge variables
\footnote{As said in B) their gauge fixings are induced by those for
primary basic gauge variables. This is a consequence of Dirac theory
of constraints \cite{a27}. Instead in numerical gravity one gives
independent gauge fixings for both the primary and secondary gauge
variables in such a way to minimize the computer time.}), the tidal
variables and the matter. Given a solution of the super-momentum and
super-Hamiltonian constraints and the Cauchy data for the tidal
variables on an initial 3-space, we can find a solution of
Einstein's equations in radar 4-coordinates adapted to a time-like
observer in the chosen gauge.
\bigskip

In Refs.\cite{a39} there is the Hamiltonian expression of radar tensors which coincide with the Riemann and Weyl tensors, expressed as radar tensors, on the solutions of Einstein equations. Moreover, by using the time-like normal to the 3-spaces and the space-like direction identified by the shift function (for solutions of Einstein equations in which it is not identically zero) it is possible to build null tetrads and to give the Hamiltonian formulation of the
Newman-Penrose formalism \cite{a40}. Therefore we get the Hamiltonian expression of Ricci and Weyl scalars and of the eigenvalues of the Weyl tensor. It is shown that the Bergmann observables \cite{a47,a48}, \cite{a46}, built with these eigenvalues cannot be DO's.

\section{The Non-Harmonic 3-Orthogonal Schwinger Time Gauges, the Post-Minkowskian
Linearization and Gravitational Waves}

In Refs.\cite{a24} the family of {\it 3-orthogonal Schwinger time
gauges} defined by the following gauge fixings ($F(\tau, \sigma^r)$
is an arbitrary numerical function)

\bea
 &&\varphi_{(a)}(\tau, \sigma^r) \approx 0, \qquad
 \alpha_{(a)}(\tau, \sigma^r) \approx 0,\nonumber \\
 &&{}\nonumber \\
 && \theta^i(\tau, \sigma^r) \approx 0,\qquad
 {}^3K(\tau, \sigma^r) \approx F(\tau, \sigma^r),
 \label{5.1}
 \eea

\noindent is defined and studied. In these gauges the instantaneous
Riemannian 3-spaces $\Sigma_{\tau}$ have a non-fixed trace ${}^3K$
of the extrinsic curvature but a diagonal 3-metric ${}^4g_{rs} = -
\sgn\, {}^3g_{rs} \approx - \sgn\, {\tilde \phi}^{2/3}\, Q_r^2\,
\delta_{rs}$ (with $Q_r = e^{\sum_{\bar a}\, \gamma_{\bar aa}\,
R_{\bar a}}$, see Eqs.(\ref{4.1})).\medskip

These gauges are {\it not harmonic gauges}. Their main property is
that in them the equations for the lapse and shift variables (see B)
of the previous Section) are {\it elliptic} PDE's inside the 3-space
like the constraints. Instead, as shown in Section V of the first
paper in Refs.\cite{a24}, the analogous equations in the family of
harmonic gauges are {\it hyperbolic} PDE's like for the tidal
variables. Therefore in harmonic gauges both the tidal variables and
the lapse and shift functions depend (in a retarded way) from the
{\it no-incoming radiation} condition on the Cauchy surface in the
past (so that the knowledge of ${}^3K$ from the initial time till
today is needed).

Instead in the family of 3-orthogonal gauges only the tidal
variables (the gravitational waves after linearization), and
therefore the 3-metric inside $\Sigma_{\tau}$, depend (in a retarded
way) on the no-incoming radiation condition. The solutions $\tilde
\phi$ and $\pi_i^{(\theta)}$ (or $\sigma_{(a)(b)}{|}_{a \not= b}$)
of the constraints and the lapse $1 + n$ and shift ${\bar n}_{(a)}$
functions depend only on the 3-space $\Sigma_{\tau}$ with fixed
$\tau$. If the matter consists of positive energy particles (with a
Grassmann regularization of the gravitational self-energies)
\cite{a24} these solutions will contain action-at-a-distance
gravitational potentials (replacing the Newton ones) and
gravito-magnetic potentials.

\bigskip

In the first paper of Ref.\cite{a24}, we studied the coupling of N
charged scalar particles plus the electro-magnetic field to ADM
tetrad gravity  in the class of asymptotically Minkowskian
space-times without super-translations. To regularize the
electro-magnetic and gravitational  self-energies both the electric
charge and the sign of the energy of the particles are
Grassmann-valued \footnote{Both quantities are two-valued. The
elementary electric charges are $Q = \pm e$, with $e$ the electron
charge. Analogously the sign of the energy of a particle is a
topological two-valued number (the two branches of the mass-shell
hyperboloid). The formal quantization of these Grassmann variables
gives two-level fermionic oscillators. At the classical level the
self-energies make the classical equations of motion ill-defined on
the world-lines of the particles. The ultraviolet and infrared
Grassmann regularization allows to cure this problem and to get
consistent solution of regularized equations of motion. See
Refs.\cite{a49} for the electro-magnetic case.}.

The introduction of the non-covariant radiation gauge (see
Ref.\cite{a21} for the special relativistic version) allows to
reformulate the theory in terms of transverse electro-magnetic
fields and to extract the generalization of the action-at-a-distance
Coulomb interaction among the particles in the non-flat Riemannian
instantaneous 3-spaces of global non-inertial frames.

\medskip

After the reformulation of the whole system in the York canonical
basis, we give the restriction of the Hamilton equations and of the
constraints to the family of {\it non-harmonic 3-orthogonal}
Schwinger time gauges.

\medskip

Then in the second paper of Ref.\cite{a24} it was shown that in this
family of non-harmonic 3-orthogonal Schwinger gauges it is possible
to define a consistent {\it linearization} of ADM canonical tetrad
gravity plus matter in the weak field approximation, to obtain a
formulation of {\it Hamiltonian Post-Minkowskian gravity with
non-flat Riemannian 3-spaces and asymptotic Minkowski background}.

\medskip

This means that the 4-metric tends to the asymptotic Minkowski
metric at spatial infinity, ${}^4g_{AB}\, \rightarrow
{}^4\eta_{AB}$. The decomposition ${}^4g_{AB} = {}^4\eta_{AB} +
{}^4h_{(1)AB}$, with a first order perturbation ${}^4h_{(1)AB}$
vanishing at spatial infinity, is only used for calculations, but
has no intrinsic meaning because the 3-spaces $\Sigma_{\tau}$ have a first order
derivation from Euclidean 3-spaces. Instead in the standard linearization one
introduces a fixed Minkowski background space-time, introduces the
decomposition $ {}^4g_{\mu\nu}(x) = {}^4\eta_{\mu\nu} +
{}^4h_{\mu\nu}(x)$ and studies the linearized equations of motion
for the small Minkowskian fields ${}^4h_{\mu\nu}(x)$. The
approximation is assumed valid over a {\it big enough characteristic
length $L$ interpretable as the reduced wavelength $\lambda / 2\pi$
of the resulting GW's} (only for distances higher of $L$ the
linearization breaks down and curved space-time effects become
relevant). See Ref.\cite{a50} for a review of all the results of the
standard approach and of the existing points of view (Damour,
Will,....) on the subject (see also Appendix A of the second paper
in Refs.\cite{a24}).
\medskip

If $\zeta << 1$ is a small a-dimensional parameter, the consistent
Hamiltonian linearization implies the following restrictions on the
variables of the York canonical basis in the family of 3-orthogonal
gauges with ${}^3K = F$ (the tidal variables $R_{\bar a}$ are slowly
varying over the length $L$ and times $L/c$; we have $({L \over
{{}^4{\cal R}}})^2 = O(\zeta)$, where ${}^4{\cal R}$ is the mean
radius of curvature of space-time)

\begin{eqnarray*}
 &&R_{\bar a}(\tau ,\vec \sigma ) = R_{(1)\bar a}(\tau,
 \vec \sigma)   = O(\zeta) << 1,\qquad
 \Pi_{\bar a}(\tau ,\vec \sigma ) = \Pi_{(1)\bar a}(\tau,
 \vec \sigma)   = {1\over {L\, G}}\, O(\zeta),\nonumber \\
 &&{}\nonumber \\
 &&\tilde \phi = \sqrt{det\, {}^3g_{rs}} = 1 + 6\, \phi_{(1)} + O(\zeta^2),\qquad
 {\bar n}_{(r)} = - \sgn\, {}^4g_{\tau r} = {\bar n}_{(1)(r)} + O(\zeta^2),\nonumber \\
 &&N = 1 + n = 1 + n_{(1)} + O(\zeta^2),\qquad \sgn\, {}^4g_{\tau\tau} =
 1 + 2\, n_{(1)} + O(\zeta^2),
 \end{eqnarray*}

 \bea
 &&{}^3K = {{12\pi\, G}\over {c^3}}\, \pi_{\tilde \phi} =
 {}^3K_{(1)} = {{12\pi\, G}\over {c^3}}\, \pi_{(1) \tilde \phi} =
 {1\over L}\, O(\zeta),\qquad
  \sigma_{(a)(b)}{|}_{a\not= b} = \sigma_{(1)(a)(b)}{|}_{a\not= b}
 = {1\over L}\, O(\zeta),\nonumber \\
 &&{}\nonumber \\
 &&{}^3g_{rs} = - \sgn\, {}^4g_{rs} = [1 + 2\, (\Gamma_r^{(1)} +
 2\, \phi_{(1)})]\, \delta_{rs} + O(\zeta^2),\qquad \Gamma^{(1)}_a =
 \sum_{\bar ar}\, \gamma_{\bar aa}\, R_{\bar a}.\nonumber \\
 &&{}
 \label{5.2}
 \eea

\medskip

The particles (whose coinciding inertial and gravitational mass is
$m_i$) are described by 3-coordinates $\eta^r_i(\tau)$ (the radar
3-coordinates of the intersection of the world-line with the
3-space: $x^{\mu}(\tau) = z^{\mu}(\tau, \eta^r_i(\tau))$) and by 3-momenta $\kappa_{ir}(\tau)$. See Refs.\cite{a24}
for the description of the electro-magnetic field. The consistency
of the Hamiltonian linearization requires the introduction of a {\it
ultra-violet cutoff $M$ for matter} such that ${{m_i}\over M}, {{
{\vec \kappa}_i}\over M} = O(\zeta)$. With similar restrictions on
the electro-magnetic field one gets that the energy-momentum tensor
of matter is $T^{AB} = T^{AB}_{(1)} + O(\zeta^2)$. This
approximation is not reliable at distances from the point particles
less than the gravitational radius $R_M = {{M\, G}\over {c^2}}
\approx 10^{25}\, M$ determined by the cutoff mass. The weak ADM
Poincar\'e generators become equal to the Poincar\'e generators of
this matter in inertial frames of Minkowski space-time plus terms of
order $O(\zeta^2)$ containing GW's and matter. Finally the GW's
described by this linearization must have wavelengths satisfying
$\lambda/2\pi \approx L >> R_M$. If all the particles are contained
in a compact set of radius $l_c$ (the source), we must have $l_c >>
R_M$ for particles with relativistic velocities and $l_c \geq R_M$
for slow particles (like in binaries).

\medskip

With this Hamiltonian linearization we can avoid to make PN
expansions, namely we get fully relativistic expressions, i.e. a PM
Hamiltonian gravity.

\bigskip

In the second paper of Refs.\cite{a24} we have found the solutions
of the super-momentum and super-Hamiltonian constraints and of the
equations for the lapse and shift functions with the Bianchi
identities satisfied. Therefore we know the first order quantities
$\pi^{(\theta)}_{(1)r}$, $\tilde \phi = 1 + 6\, \phi_{(1)}$, $1 +
n_{(1)}$, ${\bar n}_{(1)(a)}$ (the action-at-a-distance part of the
gravitational interaction) with an explicit expression for the PM
Newton and gravito-magnetic potentials. In absence of the
electro-magnetic field they are (the terms in $\Gamma^{(1)}_a =
\sum_{\bar ar}\, \gamma_{\bar aa}\, R_{\bar a}$ describe the
contribution of GW's)

\begin{eqnarray*}
 {\tilde \phi}(\tau, \vec \sigma) &=& 1 + 6\, \phi_{(1)}(\tau, \vec \sigma)
 =\nonumber \\
 &=& 1 + {{3\, G}\over {c^3}}\, \sum_i\, \eta_i\, {{\sqrt{m_i^2\, c^2 +
 {\vec \kappa}^2_i(\tau)}}\over
 { |\vec \sigma - {\vec \eta}_i(\tau)|}} -
  {3\over {8\pi}}\, \int d^3\sigma_1\, {{\sum_a\, \partial_{1a}^2\,
 \Gamma_a^{(1)}(\tau, {\vec \sigma}_1)}\over { |\vec \sigma -
 {\vec \sigma}_1|  }},\nonumber \\
 &&{}\nonumber \\
 \sgn\, {}^4g_{\tau\tau}(\tau, \vec \sigma) &=& 1 + 2\,
 n_{(1)}(\tau, \vec \sigma) =\nonumber \\
 &=& 1 -
  2\, \partial_{\tau}\, {}^3{\cal K}_{(1)}(\tau, \vec \sigma)
 - {{2\, G}\over {c^3}}\, \sum_i\, \eta_i\, {{\sqrt{m_i^2\, c^2 +
 {\vec \kappa}^2_i(\tau)}}\over { |\vec \sigma -
 {\vec \eta}_i(\tau)|}}\, \Big(1 +
 {{{\vec \kappa}^2_i}\over {m_i^2\, c^2 + {\vec \kappa}_i^2}}\Big),
 \end{eqnarray*}

 \bea
 - \sgn\, {}^4g_{\tau a}(\tau, \vec \sigma) &=& {\bar
 n}_{(1)(a)}(\tau, \vec \sigma) =
  \partial_a\, {}^3{\cal K}_{(1)}(\tau, \vec \sigma) -
 {{G}\over {c^3}}\, \sum_i\, {{\eta_i}\over {|\vec \sigma - {\vec \eta}_i(\tau)|}}\,
 \Big( \frac{7}{2}\kappa_{ia}(\tau) +\nonumber \\
 &-&\frac{1}{2} {{(\sigma^a - \eta^a_i(\tau))\, {\vec \kappa}_i(\tau) \cdot (\vec \sigma
 - {\vec \eta}_i(\tau))}\over {|\vec \sigma - {\vec \eta}_i(\tau)|^2}} \Big)
 -\nonumber \\
 &-& \int {{d^3\sigma_1}\over {4\pi\, |\vec \sigma - {\vec \sigma}_1|}}\,
 \partial_{1a}\, \partial_{\tau}\, \Big[ 2\,
 \Gamma_a^{(1)}(\tau, {\vec \sigma}_1) -
  \int d^3\sigma_2\, {{\sum_c\,  \partial_{2c}^2\,
 \Gamma_c^{(1)}(\tau, {\vec \sigma}_2)}\over {8\pi\, |{\vec \sigma}_1 -
 {\vec \sigma}_2|}}\Big],\nonumber \\
 &&{}\nonumber \\
  \sigma_{(1)(a)(b)}{|}_{a \not= b}(\tau, \vec \sigma) &=& {1\over 2}\,
 \Big(\partial_a\, {\bar n}_{(1)(b)} + \partial_b\, {\bar
 n}_{(1)(a)}\Big){|}_{a \not= b}(\tau, \vec \sigma).
 \label{5.3}
 \eea

\medskip

Then we have shown that the tidal variables $R_{\bar a}$ satisfy a
wave equation \footnote{For the tidal momenta one gets ${{8 \pi \,
G}\over {c^3}}\, \Pi_{\bar a} = \partial_{\tau}\, R_{\bar a} -
\sum_a\, \gamma_{\bar aa}\, \partial_a\, {\bar n}_{(1)(a)} +
O(\zeta^2)$.}

\bea
  \partial_{\tau}^2\, R_{\bar a}(\tau, \vec \sigma)\, &\cir& \triangle\,
 R_{\bar a}(\tau, \vec \sigma) + \sum_a\, \gamma_{\bar aa}\, \Big[
 \partial_{\tau}\, \partial_a\, {\bar n}_{(1)(a)} +\nonumber \\
 &+& \partial_a^2\, n_{(1)} + 2\, \partial_a^2\, \phi_{(1)} - 2\,
 \partial_a^2\, \Gamma_a^{(1)} + {{8\pi\, G}\over {c^3}}\,
 T_{(1)}^{aa} \Big](\tau, \vec \sigma).
 \label{5.4}
 \eea

\noindent

If we use Eqs.(\ref{5.3}) this equation becomes $(\partial^2_{\tau} - \triangle)$ \, $\sum_{\bar
b}\, M_{\bar a\bar b}\, R_{\bar a} = (known\, functional\, of\,
matter)$ with the D'Alambertian associated to the asymptotic
Minkowski 4-metric and with $M_{\bar a\bar b} = \delta_{\bar a\bar
b} - \sum_a\, \gamma_{\bar aa}\, {{\partial_a^2}\over {\triangle}}\,
\Big(2\, \gamma_{\bar ba} - {1\over 2}\, \sum_b\, \gamma_{\bar bb}\,
{{\partial_b^2}\over {\triangle}}\Big)$. The spatial operator
$M_{\bar a\bar b}$ is the main difference between the 3-orthogonal
gauges and the harmonic ones in the description of GW's.

\medskip

Therefore, by using a no-incoming radiation condition based on the
asymptotic Minkowski light-cone, we get a (complicated but tractable
due to Ref.\cite{a51}) description of gravitational waves in these
non-harmonic gauges, which can be connected to generalized
TT(transverse traceless) gauges, as {\it retarded functions of the
matter}. The results, restricted to the Solar System, are compatible
the ones of the harmonic gauges used in the BCRS frame of
Ref.\cite{a13}.\medskip

By using a Hamiltonian PM multipolar expansion in terms of Dixon
multipoles \cite{a52,a53,a54,a55} of the matter energy-momentum tensor we
get a {\it relativistic mass quadrupole emission formula}. Moreover,
notwithstanding there is no gravitational self-energy due to the
Grassmann regularization, the energy, 3-momentum and angular
momentum balance equations in GW emission are verified by {\it using
the conservation of the asymptotic ADM Poincar\'e generators} (the
same happens with the asymptotic Larmor formula of the
electro-magnetic case with Grassmann regularization as shown in Refs.\cite{a56,a57,a58,a59}).
\medskip

These GW's propagate  in non-Euclidean instantaneous 3-spaces
$\Sigma_{\tau}$ differing from the inertial asymptotic Euclidean
3-spaces at the first order (their intrinsic 3-curvature is
determined by the GW's and by the matter) and dynamically determined
by the linearized solution of Einstein equations. These 3-spaces
have a first order extrinsic curvature (with ${}^3K_{(1)}(\tau,
\sigma^r) \approx F_{(1)}(\tau, \sigma^r)$ describing the clock
synchronization convention, i.e. their shape as 3-sub-manifolds of
space-time) and a first order modification of Minkowski light-cone.
\medskip

In the third paper of Refs.\cite{a24} we eliminate the
electro-magnetic field and we evaluate all the properties of these
PM space-times:
a) the 3-volume element, the 3-distance and the intrinsic and
extrinsic 3-curvature tensors of the 3-spaces;
b) the proper time of a time-like observer;
c) the time-like and null 4-geodesics (they are relevant for the
definition of the radial velocity of stars as shown in the IAU
conventions of Ref.\cite{a60} and in study of {\it time
delays});
d) the redshift and luminosity distance. In particular we find that
the recessional velocity of a star is proportional to its luminosity
distance from the Earth at least for small distances. This is in
accord with the Hubble old redshift-distance relation which is
formalized in the Hubble law (velocity-distance relation) when the
standard cosmological model is used (see for instance
Ref.\cite{a61} on these topics). These results have a
dependence on  the non-local York time, which could play a role in
giving a different interpretation of the data from super-novae,
which are used as a support for dark energy \cite{a45}.

\medskip

With the exception of the extrinsic 3-curvature tensor all the other
quantities do not depend on the York time ${}^3K_{(1)}$ but on {\it
non-local York time} ($\triangle$ is the Laplacian associated to the
asymptotic Minkowski 4-metric)

\beq
 {}^3{\cal K}_{(1)}(\tau, \sigma^r) = \Big({1\over {\triangle}}\,
 {}^3K_{(1)}\Big)(\tau, \sigma^r).
 \label{5.5}
 \eeq

\medskip

In Subsection IIIB of the second paper in Refs.\cite{a24} it is
shown that this HPM linearization can be interpreted as the first
term of a Hamiltonian PM expansion in powers of the Newton constant
$G$ in the family of 3-orthogonal gauges. This expansion has still
to be studied.

\section{The Post-Newtonian Expansion of Post-Minkowskian Hamilton
Equations for the Particles and Dark Matter as a Relativistic
Inertial Effect due to the York Time}

We can write explicitly the linearized PM Hamilton equations for the
particles and for the electro-magnetic field: among the forces there
are both the inertial potentials and the GW's.
\medskip

In the third paper of Ref.\cite{a24} we disregarded
electro-magnetism and we studied in more detail the PM equations of
motion of the particles. If we use Eqs.(\ref{5.3}) and the retarded
solution of Eqs.(\ref{5.4}) for the GW's, the regularized equations
of motion depend only on the particles and have the form ${\ddot
{\vec \eta}}_i(\tau) = {1\over {m_i}}\, {\vec F}_i(\tau )$ with the
forces depending on the positions and velocities of all the
particles. Eqs.(\ref{5.3}) imply that the equation for particle "i"
is independent from $m_i$ (equality of inertial and gravitational
masses). The effective force ${\vec F}_i(\tau)$ contains\medskip

a) the contribution of the lapse function ${\check n}_{(1)}$, which
generalizes the Newton force;\medskip

b) the contribution of the shift functions ${\check {\bar
n}}_{(1)(r)}$, which gives the gravito-magnetic effects;\medskip

c) the retarded contribution of GW's, described by the functions
$\Gamma_r^{(1)}$;\medskip

d) the contribution of the volume element $\phi_{(1)}$ (${\tilde
\phi} = 1 + 6\, \phi_{(1)} + O(\zeta^2)$), always summed to the
GW's, giving forces of Newton type;\medskip

e) the contribution of the inertial gauge variable (the non-local
York time) ${}^3{\cal K}_{(1)} = {1\over {\triangle}}\,
{}^3K_{(1)}$.\medskip

While in the electro-magnetic case in SR \cite{a49} the
regularized equations of motion of the particles obtained by using
the  Lienard-Wiechert solutions for the electro-magnetic field are
independent by the type of Green function (retarded or advanced or
symmetric) used, this is not strictly true in the gravitational
case. The effect of retardation is only pushed to $O(\zeta^2)$ and
should appear in a study of the second order equations of motion.
\bigskip

Then we studied the Post-Newtonian (PN) expansion of these
regularized PM equations of motion for the particles. We found  that
the particle 3-coordinates $\eta^r_i(\tau = ct) = {\tilde
\eta}_i^r(t)$ satisfy the equation of motion

\bea
 m_i\, {{d^2\, {\tilde \eta}_i^r(t)}\over {dt^2}}
 &=& m_i\, \Big[-G\, {{\partial}\over {\partial\,
 {\tilde \eta}_i^r}}\,  \sum_{j \not= i}\,  {{m_j}\over
 {|{\vec {\tilde \eta}}_i(t) - {\vec {\tilde \eta}}_j(t)|}} -
  {1\over c}\, {{d {\tilde \eta}^r_i(t)}\over {dt}} \Big(
  \partial^2_t{|}_{{\vec {\tilde \eta}}_i}\, {}^3{\tilde
  {\cal K}}_{(1)} +\nonumber \\
  &+& 2\, \sum_u\, v_i^u(t)\, {{\partial\, \partial_t{|}_{{\vec {\tilde \eta}}_i}\,
 {}^3{\tilde {\cal K}}_{(1)} }\over {\partial\, {\tilde \eta}_i^u}}
 + \sum_{uv}\, v_i^u(t)\, v^v_i(t)\, {{\partial^2\, {}^3{\tilde
 {\cal K}}_{(1)}}\over {\partial\, {\tilde \eta}_i^u\, \partial\, {\tilde \eta}_i^v}}
 \Big) (t, {\vec {\tilde \eta}}_i(t)) \Big] +\nonumber \\
 &+& F^r_{(1PN)}(t) + (higher\, PN\, orders),
 \label{6.1}
 \eea

\noindent where at the lowest order we find the standard Newton
gravitational force ${\vec F}_{i (Newton)}(t) = - m_i\, G\,
{{\partial}\over {\partial\, {\tilde \eta}_i^r}}\,  \sum_{j \not=
i}\,  {{m_j}\over {|{\vec {\tilde \eta}}_i(t) - {\vec {\tilde
\eta}}_j(t)|}}$.\medskip

If we put ${}^3K_{(1)} = 0$, the forces ${\vec F}_{Newton} + {\vec
F}_{(1PN)}$ reproduce the standard results about binaries found with
a different type of approximation in Ref.\cite{a62}.
\medskip

Therefore the (arbitrary in these gauges) double rate of change in
time of the trace of the extrinsic curvature creates a 0.5 PN
damping (or anti-damping since the sign of the inertial gauge
variable ${}^3{\cal K}_{(1)}$ of Eq.(\ref{5.5}) is not fixed) effect
on the motion of particles. This is an {\it inertial effect} (hidden in the
lapse function) not existing in Newton theory where the Euclidean
3-space is absolute.\medskip

In the 2-body case we get that for Keplerian circular orbits of
radius $r$ the modulus of the relative 3-velocity can be written in
the form $\sqrt{{{G\, (m + \triangle\, m(r))}\over r}}$ with
$\triangle\, m(r)$ function only of ${}^3{\cal K}_{(1)}$.

Now the {\it rotation curves of spiral galaxies} (see Refs.\cite{a63,a64,a65}
for  reviews) imply that the relative 3-velocity goes to constant
for large $r$ (instead of vanishing like in Kepler theory).This
result can be simulated by fitting $\triangle\, m(r)$ (i.e. the
non-local York time) to the experimental data:  as a consequence
$\triangle\, m(r)$ is interpreted as a {\it dark matter halo} around
the galaxy. With our approach this dark matter would be a {\it
relativistic inertial effect} consequence of the a non-trivial shape
of the non-Euclidean 3-space as a 3-sub-manifold of space-time.
\medskip

We find that  Eq.(\ref{6.1}) can be rewritten in the form

 \bea
 \frac{d}{dt}\Big[ \,m_i\Big(1+\frac{1}{c}\,\frac{d}{dt}\,{}^3{\tilde {\cal K}}_{(1)}(t,
 {\vec {\tilde \eta}}_i(t))\,\Big)\, {{d\, {\tilde \eta}_i^r(t)}\over {dt}}\Big]
 &\cir &\,-G\, {{\partial}\over {\partial\,
 {\tilde \eta}_i^r}}\,  \sum_{j \not= i}\, \eta_j\, {{m_i\,m_j}\over
 {|{\vec {\tilde \eta}}_i(t) - {\vec {\tilde \eta}}_j(t)|}}+\nonumber\\
 &&\nonumber\\
 &+&{\cal O}(\zeta^2).
 \label{6.2}
\eea

We see that the term in the non-local York time can be {\it
interpreted} as the introduction of an {\it effective (time-,
velocity- and position-dependent) inertial mass term} for the
kinetic energy of each particle: $m_i\, \mapsto\,
m_i\,\Big(1+\frac{1}{c}\,\frac{d}{dt}\,{}^3{\tilde {\cal
K}}_{(1)}(t, {\vec {\tilde \eta}}_i(t))\,\Big) $ in each
instantaneous 3-space. Instead in the Newton potential there are the
gravitational masses of the particles, equal to the inertial ones in
the 4-dimensional space-time due to the equivalence principle.
Therefore the effect is due to a modification of the effective
inertial mass in each non-Euclidean 3-space depending on its shape
as a 3-sub-manifold of space-time: {\it it is the equality of the
inertial and gravitational masses of Newtonian gravity to be
violated}! In Galilei space-time the Euclidean 3-space is an
absolute time-independent notion like Newtonian time: the
non-relativistic non-inertial frames live in this absolute 3-space
differently from what happens in SR and GR, where they are (in
general non-Euclidean) 3-sub-manifolds of the space-time.

\medskip

A similar interpretation can be given for the other two  main
signatures of the existence of dark matter in the observed masses of
galaxies and clusters of galaxies, namely the virial theorem
\cite{a66,a67,a68} and weak gravitational lensing \cite{a69}
\cite{a67,a70}.
\medskip

This option for explaining (at least part of) dark matter differs:\medskip

1) from the non-relativistic MOND approach \cite{a71} (where one
modifies Newton equations);\medskip

2) from modified gravity theories like the $f(R)$ ones (see for
instance Refs.\cite{a72}; here one gets a modification of the Newton
potential);\medskip

3) from postulating the existence of WIMP particles \cite{a73}.
\medskip

Let us also remark that the 0.5PN effect has origin in the lapse
function and not in the shift one, as in the gravito-magnetic
elimination of dark matter proposed in Ref.\cite{a74}.

\medskip

As a consequence of the property of non-Euclidean  3-spaces of being
3-sub-manifolds of Einstein space-times (independently from
cosmological assumptions) there is the possibility of describing
part (or maybe all) dark matter as a {\it relativistic inertial
effect}. As we have seen the three main experimental signatures of
dark matter can be explained in terms of the non-local York time
${}^3{\cal K}_{(1)}(\tau, \vec \sigma)$, the inertial gauge variable
describing the general relativistic remnant of the gauge freedom in
clock synchronization.

\section{Clock Synchronization and Relativistic Metrology}

Since in GR the gauge freedom is the
arbitrariness in the choice of the 4-coordinates,  a similar
arbitrariness is expected in the non-inertial frames of SR and this is described in Refs. \cite{a21,a22}.

However, at the experimental level the description of  matter (and
also of the spectra of light from stars)  is not based on DO's or 4-scalars but is {\it intrinsically
coordinate-dependent}, namely is connected with the {\it
metrological conventions} used by physicists, engineers and
astronomers for the modeling of space-time (see Ref.\cite{a75} for a review on relativistic metrology). The basic conventions
are\medskip

a) An atomic clock as a standard of time;\medskip

b) The 2-way velocity of light in place of a standard of
length;\medskip

c) A conventional reference frame centered on a given observer as a
standard of space-time (GPS is an example of such a
standard);\medskip

\noindent and the adopted astronomical reference frames are:\medskip

A) The description of satellites around the Earth is done by means
of NASA coordinates \cite{a41} either in ITRS (the terrestrial frame
fixed on the Earth surface)\cite{a42} or in GCRS (the geocentric
frame centered on the Earth center) (see Ref.\cite{a43}).
\medskip

B) The description of planets and other objects in the Solar System
uses BCRS (a barycenter quasi-inertial Minkowski frame, if
perturbations from the Milky Way are ignored \footnote{Essentially
it is defined as a {\it quasi-inertial system}, {\it non-rotating}
with respect to some selected fixed stars, in Minkowski space-time
with nearly-Euclidean Newton 3-spaces. The qualification {\it
quasi-inertial} is introduced  to take into account GR, where
inertial frames exist only locally. More exactly it is a PM Einstein
space-time with 3-spaces having a very small extrinsic curvature of
order $c^{-2}$ and with a PN treatment of the gravitational field of
the Sun and of the planets in a special harmonic gauge of Einstein
GR (see Ref.\cite{a76} for possible gravitational anomalies inside
the Solar System).}), centered in the barycenter of the Solar
System, and ephemerides (see Ref.\cite{a43}).
\medskip

C) In astronomy the positions of stars and galaxies are determined
from the data (luminosity, light spectrum, angles) on the sky as
living in a 4-dimensional nearly-Galilei space-time with the
celestial ICRS \cite{a45} frame  considered as a "quasi-inertial
frame" (all galactic dynamics is Newtonian gravity), in accord with
the assumed validity of the cosmological and Copernican principles.
Namely one assumes a homogeneous and isotropic cosmological
Friedmann-Robertson - Walker solution of Einstein equations (the
standard $\Lambda$CDM cosmological model). In it the constant
intrinsic 3-curvature of instantaneous 3-spaces is  nearly zero as
implied by the CMB data \cite{a45}, so that Euclidean 3-spaces (and
Newtonian gravity) can be used. However, to reconcile all the data
with this 4-dimensional reconstruction one must postulate the
existence of dark matter and dark energy as the dominant components
of the classical universe after the recombination 3-surface!

What is still lacking is a PM extension of the celestial frame such
that the PM BCRS frame is its restriction to the solar system inside
our galaxy. Hopefully this will be achieved with the ESA GAIA
mission devoted to the cartography of the Milky Way \cite{a77}.

\medskip

The recombination 3-surface is the natural Cauchy surface for using
classical GR in the description of the 3-universe after the end of
the preceding quantum phases of its evolution (it is a kind of
Heisenberg cut between quantum cosmology and classical
astrophysics). Let us also remark that the fixed stars of star
catalogues \cite{a44} may be considered as a phenomenological
definition of {\it spatial infinity} in asymptotically Minkowskian
space-times: their spatial axes define an asymptotic inertial frame
centered on an asymptotic inertial observer.

\medskip

Therefore, in every generally covariant theory of gravity  the
freedom in the choice of the gauge, i.e. of the 4-coordinate system
of space-time and of the time-like observer origin of the
3-coordinates, disappears when we want to make comparison with
experimental data: we must choose those mathematical gauges which
are compatible with the metrological conventions.

\bigskip

Since, as said in the Introduction, at the experimental level {\it
the description of matter is intrinsically coordinate-dependent},
namely is connected with the conventions used by physicists,
engineers and astronomers for the modeling  of space-time, we have
to choose a gauge (i.e. a 4-coordinate system) in non-modified
Einstein gravity which is in agreement with the observational
conventions in astronomy.\medskip

Since ICRS \cite{a44} has diagonal 3-metric, our 3-orthogonal gauges
are a good choice. We are left with the inertial gauge variable
${}^3{\cal K}_{(1)} = {1\over {\triangle}}\, {}^3K_{(1)}$ not
existing in Newtonian gravity. As already said the suggestion is to
try to fix ${}^3{\cal K}_{(1)}$ in such a way to eliminate dark
matter as much as possible.\medskip

The open problem is the determination of the non-local York time
from the data. From what is known from the Solar System and from
inside the Milky Way near the galactic plane, it seems that it is
negligible near the stars inside a galaxy. On the other hand, it is
non zero near galaxies and clusters of galaxies of big mass. However
only a mean value in time of time- and space-derivatives of the
non-local York time can be extracted from the data. At this stage it
seems that the non-local York time is relevant around the galaxies
and the clusters of galaxies where there are big concentrations of
mass and the dark matter haloes and that it becomes negligible
inside the galaxies where there is a lower concentration of mass.
Instead there is no indication on its value in the voids existing
among the clusters of galaxies.

\medskip

However if we do not {\it know the non-local York time on all the
3-universe} at a given $\tau$ we cannot get an experimental
determination of the York time ${}^3K_{(1)}(\tau, \vec \sigma) =
\triangle\, {}^3{\cal K}_{(1)}(\tau, \vec \sigma)$. Therefore some
phenomenological parametrizations of ${}^3{\cal K}_{(1)}(\tau, \vec
\sigma)$ will have to be devised to see the implications for
${}^3K_{(1)}(\tau, \vec \sigma)$. As said in the Introduction, a
phenomenological determination of the York time would help in trying
to get a PM extension of the existing quasi-inertial Galilei
Celestial reference frame (ICRS). Then automatically BCRS would be
its quasi-Minkowskian approximation for the Solar System. Let us
remark that the 3-spaces can be quasi-Euclidean (i.e. with a small
3-curvature tensor), as required by CMB data \cite{a45} in the
astrophysical context, even when their shape as 3-sub-manifolds of
space-time is not trivial and is described by a not-small York time.
\medskip

This would be the way out from the gauge problem in general
relativity: the observational conventions for matter would select a
reference system of 4-coordinates for PM space-times in the
associated 3-orthogonal gauge.

\section{Conclusions}

I have given a full review of an approach  to
asymptotically Minkowskian classical canonical GR based on a
description of global non-inertial frames centered on a time-like
observer which is suggested by relativistic metrology. The gauge
freedom in clock synchronization, which does not exist in Galilei
space-time (Newton time is absolute) and is not restricted in
Minkowski space-time (the whole class of admissible 3+1 splittings in this
absolute space-time is allowed), is restricted to the gauge freedom connected
with the inertial gauge variable ${}^3K$, the York time, which
determines the shape of the instantaneous (in general non-Euclidean)
3-spaces as 3-sub-manifolds of the space-time.
\medskip

The study of canonical ADM tetrad gravity in asymptotically
Minkowskian space-times in the York canonical basis allowed to find
the family of non-harmonic 3-orthogonal Schwinger time gauges and to
define a HPM linearization in them. The main properties of these
non-harmonic gauges are that only the GW's (but not the lapse and
shift functions) are retarded quantities with a no-incoming
radiation condition and that one can naturally find which quantities
depend upon the York time.

\medskip

I have described relativistic particle mechanics  in  GR.
The more surprising result is that in the PN expansion of the PM
equations of motion there is a 0.5PN term in the forces depending
upon the York time. This opens the possibility to describe dark
matter as a relativistic inertial effect implying that the effective
inertial mass of particles in the 3-spaces is bigger of the
gravitational mass because it depends on the York time (i.e. on the
shape of the 3-space as a 3-sub-manifold of the space-time: this is
impossible in Newton gravity in Galilei space-time and leads to a
violation of the Newtonian equivalence principle).

\bigskip

At a more mathematical level some open problems under investigation
are:\medskip

A) The quantization of the massive Klein-Gordon field in
non-inertial frames of Minkowski space-time. Is it possible to evade
the no-go theorem of Ref.\cite{a78} and to get unitary
evolution? And to extend to GR?\medskip

B) Find the second order of the HPM expansion to see whether in PM
space-times there is the emergence of hereditary terms (see
Refs.\cite{a50,a79}) like the ones present in harmonic gauges. Like
in standard approaches (see the review in Appendix A of the second
paper in Refs.\cite{a24}) regularization problems may arise at the
higher orders.\medskip

C) Study  the PM equations of motion of the transverse
electro-magnetic field trying to find Lienard-Wiechert-type
solutions (see Subsection VB of the second paper in
Refs.\cite{a24}). Study astrophysical problems where the
electro-magnetic field is relevant.\medskip

D) Try to find the final true DO's of GR by using approximate solutions of the super-Hamiltonian
and super-momentum constraints. This would allow to make a multi-temporal
quantization (see Ref. \cite{a80}) of the approximate solution (for instance the linearized
HPM theory over the asymptotic Minkowski space-time as shown in the second paper of Ref.\cite{a39}), in which only
the tidal DO variables are quantized but not the inertial gauge ones.
In this way the space-time would remain a 4-manifold with the gauges determined by relativistic metrology:
only the eigenvalues of the 3-metric of the non-Euclidean 3-spaces would be quantized with an
induced quantization of 3-lengths, 3-areas and 3-volumes
to be compared with the results of loop quantum gravity.

\medskip

E) Find the canonical transformation from  the York canonical basis
to the  Ashtekar variables \cite{a81,a82}, \cite{a4,a12}, in asymptotically Minkowskian
space-times.

\bigskip

Instead at the physical level the next big challenge after dark
matter is {\it dark energy} in cosmology \cite{a45} (see
Ref.\cite{a83,a84} for what we really know). Even if in cosmology we
cannot use canonical gravity, in the first paper of Ref.\cite{a24}
it is shown that the usual non-Hamiltonian 12 ADM equations can be
put in a form allowing to use the interpretations based on the York
canonical basis by means of the expansion and the shear of the
Eulerian observers.\medskip

Let us remark that in the Friedmann-Robertson-Walker (FRW)
cosmological solution the Killing symmetries connected with
homogeneity and isotropy imply ($\tau$ is the cosmic time, $a(\tau)$
the scale factor) ${}^3K (\tau) = - {{\dot a(\tau)}\over {a(\tau)}}
= - H$, namely the York time is no more a gauge variable but
coincides with the Hubble constant. However at the first order in
cosmological perturbations we have ${}^3K = - H + {}^3K_{(1)}$ with
${}^3K_{(1)}$ being again an inertial gauge variable. Instead in
inhomogeneous space-times without Killing symmetries like the
Szekeres ones \cite{a85,a86,a87} the York time remains an inertial gauge
variable.
\medskip

Also in the back-reaction approach \cite{a88,a89,a90} (see also
Ref.\cite{a91}) to dark energy, according to which dark energy is a
byproduct of the non-linearities of general relativity when one
considers spatial mean values on large scales to get a cosmological
description of the universe taking into account the inhomogeneity of
the observed universe, one gets that the spatial average of the York
time (a 3-scalar gauge variable) gives the effective Hubble constant
of that approach.
\medskip

Therefore the York time has a central position also in the main
quantities on which relies the interpretation of dark energy in the
standard $\Lambda$CDM cosmological model (Hubble constant, the old
Hubble redshift-distance relation  replaced in FRW cosmology with
the velocity distance relation or Hubble law). As a consequence it
looks reasonable to investigate on a possible gauge origin also of
dark matter.

\medskip

As a first step we have considered a perfect fluid as matter in the
first order of HPM expansion \cite{a92} adapting to tetrad gravity
the special relativistic results of Refs.\cite{a93,a94} (based on the
approach of Ref.\cite{a95}). Since in our formalism all the
canonical variables in the York canonical basis, except the angles
$\theta^i$, are 3-scalars, we can complete Buchert's formulation of
back-reaction \cite{a88} by taking the spatial average of nearly all
the PM Hamilton equations in our non-harmonic 3-orthogonal gauges.
This will allow to make the transition from the PM space-time
4-metric to an inhomogeneous cosmological one (only conformally
related to Minkowski space-time at spatial infinity) and to try to
reinterpret also the dark energy as a relativistic inertial effect
(and not only as a non-linear effect of inhomogeneities). The role
of the York time, now considered as an inertial gauge variable, in
the theory of back-reaction  and in the identification of what is
called dark energy \footnote{As we have already said at the PM level
the red-shift and the luminosity distance depend upon the York time,
and this could play a role in the interpretation of the data from
super-novae.} is completely unexplored.\medskip

Moreover the recent point of view of Ref.\cite{a96} taking into
account the relevance of the voids among the clusters of galaxies
suggests to try to develop a phenomenological parametrization of the
York time to see whether we can simultaneously fit the data on dark
matter and make contact with the back-reaction approach to dark
energy.\medskip

Finally let us remark that in Eq.(\ref{4.5}) we showed that in the
York canonical basis the York time contributes with a negative term
to the kinetic energy in the ADM energy. It would also play a role
in a study to be done on the reformulation of the Landau-Lifschitz
energy-momentum pseudo-tensor as the energy-momentum tensor of a
viscous pseudo-fluid. It could be possible that for certain choices
of the York time the resulting effective equation of state has
negative pressure, realizing also in this way a simulation of dark
energy.

\medskip

Is it possible to find a 3-orthogonal gauge in a inhomogeneous
Einstein space-time in which the inertial gauge variable York time
allows to eliminate both dark matter and dark energy through the
choice of a 4-coordinate system to be used as a consistent PM ICRS
saving the main good properties of the standard $\Lambda$CDM model?

\bigskip

As a final remark let us notice that
when there is a perfect fluid with unit time-like 4-velocity
$U^A(\tau, \vec \sigma)$, there is also the congruence of its
time-like flux curves: in general it is not surface-forming and it
is independent from the previous two congruences. If $\Big(U^A(\tau,
\vec \sigma); {\cal U}^A_{(a)}(\tau, \vec \sigma)\Big)$ is an
ortho-normal tetrad carried by a flux line, the connection of these
4-vectors to the ortho-normal tetrad of the Eulerian observers  is

\bea
 U^A(\tau, \vec \sigma) &=& \Gamma\, \Big( l^A + \sum_a\, \beta_{(a)}\,
 {}^4{\buildrel \circ \over {\bar E}}^A_{(a)}\Big)(\tau, \vec
 \sigma),\nonumber \\
 {\cal U}^A_{(a)}(\tau, \vec \sigma) &=& \Big(t_{(a)}\, l^A + \sum_b\,
 \gamma_{(a)(b)}\, {}^4{\buildrel \circ \over
 {\bar E}}^A_{(b)}\Big)(\tau, \vec \sigma),
 \label{8.1}
 \eea

\noindent with $t_{(a)}(\tau, \vec \sigma) = \Big( \sum_b\,
\gamma_{(a)(b)}\, \beta_{(b)}\Big)(\tau, \vec \sigma)$ and
$\Big[\sum_{cd}\, \Big(\delta_{(c)(d)} - \beta_{(c)}\,
\beta_{(d)}\Big)\, \gamma_{(a)(c)}\, \gamma_{(b)(d)}\Big](\tau, \vec
\sigma) = \delta_{(a)(b)}$. When the vorticity of the fluid
vanishes, so that its 4-velocity is surface forming, there is a 3+1
splitting of space-time determined by the irrotational fluid. While
in SR we can always choose a global non-inertial frame coinciding
with these 3+1 splitting, in GR we have to show that there is a
gauge fixing on the inertial gauge variable ${}^3K(\tau, \vec
\sigma)$ (the York time) allowing this identification. These
problems are studied in Ref.\cite{a92}.

\bigskip

Let us remark that Eqs.(\ref{8.1}) establish the bridge between our
3+1 point of view and the 1+3 point of view of Refs.\cite{a97,a98,a99},
where one describes both the gravitational field and the matter as
seen by a generic family of observers with 4-velocity $U^A(\tau,
\vec \sigma)$. Most of the results in cosmology (see for instance
Refs.\cite{a100,a101}) are presented in the 1+3 framework. However, in the
1+3 point of view vorticity is an obstruction to formulate the
Cauchy problem (3-spaces are not existing; each observer uses as
rest frame the tangent 3-space orthogonal to the 4-velocity) and
there is no natural way to identify the inertial gravitational gauge
variables of the Hamiltonian formalism based on Dirac's constraint
theory (see also Appendix A of the first paper in Refs.\cite{a24}).

\end{document}